\documentclass[trackchanges]{aastex701}

\newcommand{\SH}[1]{\textcolor{blue}{#1}}

\begin{document}

\title{High-energy Multi-messenger Emission from Galaxy Clusters in the Local Universe }

\author[0000-0002-0458-0490]{Saqib Hussain}
\affiliation{University of Nova Gorica, Vipavska 13,SI-5000, Nova Gorica,
Slovenia.}
\email[show]{saqib.hussain@ung.si}

\author[orcid=0000-0001-8484-7791]{	Gabrijela Zaharijas} 
\affiliation{University of Nova Gorica, Vipavska 13,SI-5000, Nova Gorica,
Slovenia.}
\email{gabrijela.zaharijas@ung.si}

\author[]{Klaus Dolag}
\affiliation{University Observatory Munich, Scheinerstr 1, D-81679 Munchen, Germany.}
\affiliation{Max Planck Institute for Astrophysics, Karl-Schwarzschild-Str 1, D-85741 Garching, Germany.}
\email{dolag@usm.lmu.de}



\begin{abstract}
The origin of diffuse neutrinos and $\gamma-$rays is unknown, and galaxy clusters hosting AGN and starburst galaxies are the most probable sources of these cosmic messengers.
%
In this work, we investigate the diffuse $\gamma-$ray and neutrino emission from the Virgo, Perseus, and Coma clusters using a detailed numerical method, combining MHD simulations with Monte Carlo methods.
The MHD simulation provides the distributions of temperature, gas, and magnetic field in clusters. The Monte Carlo simulations are used to investigate the cosmic-ray (CR) propagation in ICM and subsequently the secondaries stemming from CRs.
Our primary assumption is that CR injection follows the gas density of clusters, providing a physically motivated approximation.
High-density regions in clusters are associated with strong turbulence and prominent shock structures, making them the natural sites for efficient CR acceleration.
Our predicted $\gamma-$ray flux from the individual clusters lies well below the present LHAASO upper limits. 
The MAGIC observations of the central source NGC~$1275$ of the Perseus cluster are significantly higher than our results.
Further, we estimated the cumulative $\gamma$-ray and neutrino fluxes from clusters with masses $\gtrsim 5\times 10^{13}\, M_{\odot}$ in the local Universe (within $500$~Mpc).
The diffuse $\gamma-$ray flux reported by the Fermi-LAT collaboration is significantly higher than our results.
Our predictions are consistent with IceCube’s existing upper limits on the unresolved neutrino flux from galaxy clusters ($M > 10^{14}\, M_{\odot}$) up to $z = 2$.
%
\end{abstract}



\section{Introduction}\label{sec:intro}
The origin of the diffuse neutrino and $\gamma-$ray background remains one of the major unsolved problems in astrophysics. This diffuse emission is defined as the emission remaining after subtracting the contributions from all individually resolved astrophysical sources. Studying it offers a
unique probe of the high-energy Universe.
The inherent link of diffuse $\gamma-$rays and neutrinos with CRs highlights the importance of investigating the most powerful accelerators in the Universe.
Several studies \citep[see e.g.,][]{ajello2015origin, Raniere2022isotropic} (and references therein) explored the origin of diffuse $\gamma-$rays, predicting that a maximum of $40\,\%$ of the observed flux by Fermi-LAT can be associated with AGNs.
Moreover, associating the diffuse neutrinos observed by the IceCube with an astrophysical source class is even more complex.

%

%
The comparable magnitude of diffuse $\gamma-$rays, neutrinos and CRs fluxes suggests a possibility of a common origin, potentially arising from a single class of astrophysical sources \citep{ahlers2018opening,fang2018linking,hussain2023diffuse}. 
Galaxy clusters hosting AGNs are one of the most suitable candidates to produce high-energy multi-messenger signals due to their dense environment filled with hot gas ($\sim 10^{8}$~K) and strong magnetic field ($\sim 1\,\mu \mathrm{G}$).
Massive galaxy clusters are formed through hierarchical merging of smaller structures, including galaxies and groups.
Galaxy clusters comprise two primary mass components: dark matter, which interacts only gravitationally, and gas, which is subject to pressure forces and energy dissipation. These processes give rise to shocks and turbulence in the ICM during mergers of galaxies and groups.
These mergers are among the most violent processes in the universe that can release a large amount of energy ($\sim 10^{64}$~erg), a fraction of which derives shocks and turbulence capable of accelerating CRs to very high energy $\gtrsim 10^{18}$~eV \citep{abbasi2014indications, kim2019filaments}.
In addition, the gravitational potential energy of infalling galaxies and groups dissipates, heating the gas of ICM to very high temperature $\sim 10^{8}$~K \citep{muldrew2015protoclusters, molnar2016cluster}. 
%

%
Extended radio structures known as giant radio halos have been observed in a fraction of galaxy clusters \citep{van2019diffuse}, providing strong evidence for nonthermal phenomena associated with relativistic electrons and high-energy CRs \citep{brunetti2014cosmic, brunetti2017relativistic,adam2021gamma,kushnir2024coma}.
Clusters can trap these energetic particles because of their large size ($\sim1$~Mpc) and strong magnetic field ($\sim \mu \rm{G}$).
These conditions suggest that clusters can produce high-energy $\gamma-$rays and neutrinos, especially through proton-proton (pp) collisions and pion decay.
In addition, inverse-Compton scattering of cosmic microwave background (CMB) photons by secondary electrons and positrons produced in hadronic interactions in the ICM can contribute to $\gamma-$ray emission in the GeV–TeV energy range \citep{boss2025simulating}.



Several studies \cite[][see also references therein]{murase2013testing,fang2016high, nishiwaki2021particle, nishiwaki2023high}  predicted that clusters can contribute to a significant percentage of diffuse neutrino and $\gamma-$ray backgrounds.
In \cite{hussain2021high, hussain2023diffuse},  the authors used a detailed numerical approach to estimate the contribution of clusters across redshifts ($z\leq5.0$) to diffuse backgrounds.
Their results indicate a substantial cluster contribution to the diffuse neutrino background above $100$~ TeV energy and to the gamma rays above $10$~ GeV energy.

In the local Universe, individual clusters such as Coma, Perseus, and Virgo are among the promising sources of neutrinos and $\gamma-$rays.
Using analytical or semi-analytical approaches, many studies \citep{brunetti2017relativistic, nishiwaki2021particle,nishiwaki2023high, nishiwaki2024low} have estimated the $\gamma-$ray and neutrino emissions from Coma-like clusters, predicting that diffuse signals could be detected by future observatories. 
These studies used simplified modeling, for instance, considering spherically symmetric profiles for density and magnetic field in clusters. 
Previous studies \citep{dolag2005constrained, hussain2021high} demonstrated that the properties of the clusters deviate significantly from spherical symmetry, which in turn affects the resulting emission patterns of $\gamma-$rays and neutrinos.
Moreover, all of the above-mentioned studies assumed the injection of CRs only by internal sources in the cluster core, disregarding the fact that accretion shocks at the virial radius can also accelerate CRs to very high energies \citep{ilani2024galaxy}.

Using $15$ years of Fermi-LAT data, a systematic study was carried out on $300$ massive galaxy clusters identified in $2500\, \mathrm{deg}^{2}$ SPT-SZ survey \citep{bleem2015galaxy}, but did not detect significant $\gamma-$ray emission from these sources \citep{manna2024search}.
MAGIC collaboration observed the Perseus cluster region and reported the detection of high-energy $\gamma-$rays from the central source NGC1275, but did not detect diffuse emission \citep{ahnen2016deep}.
Recently, the Large High Altitude Air Shower Observatory (LHAASO) collaboration studied the diffuse $\gamma-$ray emissions from Virgo, Perseus, and Coma clusters, but no significant emission was detected, providing upper-limits for these sources.
The CTA collaboration will focus on the Perseus cluster, exploring whether it can emit $\gamma-$rays \citep{cta2018science}. IceCube-Gen2 and KM3Net would be able to detect neutrinos from clusters. Future detections from these experiments may establish galaxy clusters as a new class of sources that can emit high-energy messengers.

Minding the limitation mentioned above and the prospect for near-term future observations motivate us to explore the multi-messenger emission from galaxy clusters.
In this work, our focus is on individual galaxy clusters such as Virgo, Perseus, and Coma in the local Universe, the most prominent ones that can emit high-energy cosmic messengers.
We constrain the diffuse emission, like the emission produced by the large-scale CR interactions in the ICM.
We use a rigorous numerical method, combining MHD simulation with Monte Carlo methods to study the CR propagation in the ICM and investigate the production of secondaries, including $\gamma-$rays and neutrinos.

This article is arranged as follows: The sec. \ref{sec:method} explain the methodology of this work and in sec. \ref{sec:results} we presented our findings. The results are  discussed in sec. \ref{sec:discussion} and in the last section we presented then summary of this work.

\section{Methods}\label{sec:method}

Diffuse emission comes from the large-scale CR interactions in the ICM, whereas the point-source emission by the embedded AGN or starburst galaxies in clusters is compact and localized. However, both components can produce similar spectral signatures.
Massive galaxy clusters such as Virgo and Perseus host single prominent central sources, M87 and NGC1275, respectively.
Clusters can have multiple central sources, such as the Coma cluster that hosts NGC4874 and NGC4889.  
These central sources can inject high-energy CRs into the ICM \citep{hardcastle2020radio}.
This scenario refers to the internal sources at the center of clusters.
In addition, cosmological accretion shocks formed around the virial radius of clusters can accelerate CR and inject them in ICM \citep{ilani2024galaxy, keshet2024radio}.
Therefore, $\gamma-$ray and neutrino fluxes of clusters can have two components, central source emission and diffuse emission; establishing clear distinctions between them is rather complex, particularly given the limited angular resolution of high-energy observations.

%
%
%
%

%
Modeling the diffuse emission from galaxy clusters requires several assumptions concerning the properties of CRs. These include the injection spectrum, such as the spectral index and the cut-off energy. Additionally, the spatial distribution of CRs within the cluster structure must be specified, for example, whether they are concentrated near the central regions or spread throughout the ICM. The CR composition is another crucial factor that involves the relative abundance of protons and heavier nuclei, as well as the potential spatial variations of this composition. Furthermore, the confinement and transport of CRs in clusters are governed by the strength and configuration of the magnetic field, which determine their diffusion, streaming, and eventual escape.
Importantly, all of these are poorly constrained, as they are not directly measurable. These parameters can be inferred indirectly through observations and modeling.

%

To reliably model the properties of the turbulent environment of clusters across redshifts, MHD simulations offer one of the most robust methods.
Concerning CR transport in the ICM, Monte Carlo simulations are the most effective approach for studying CR transport in the ICM, accounting for all relevant CR interactions and energy-loss mechanisms. 

In our previous studies \citep{hussain2021high, hussain2023diffuse, hussain2024neutrinos}, we develop a modeling framework that combined constrained MHD simulations \citep{dolag2005constrained} with Monte Carlo simulations to model the resulting 
$\gamma-$ray and neutrino emission. 
These MHD simulations were primarily designed to model the large-scale magnetic field and therefore do not fully capture the detailed thermodynamical structure of galaxy clusters.
Moreover, in these studies, CRs were injected only at the cluster centers, whereas CR acceleration can also occur in cluster outskirts, for example at merger-driven shocks and through large-scale turbulence.


In this work, we instead employ the MHD simulation known as Simulating the LOcal Web (SLOW) \footnote{https://www.usm.uni-muenchen.de/~dolag/Simulations/} \citep{dolag2023simulating}, which combine observationally constrained initial conditions with modern hydrodynamical and MHD modeling, providing a self-consistent and high-resolution description of the 3-dimensional (3D) gas and magnetic field distributions within galaxy clusters. This makes the SLOW simulations significantly better suited for accurately characterizing realistic cluster environments, including their non-spherical morphology as well as localized anomalies and substructures arising from mergers and dynamical activity. Furthermore, for CR injection we adopt a spatially extended model that follows the cluster gas density, as described below, allowing for CR acceleration beyond the cluster core.
In this paper, we employ state-of-the-art MHD simulations together with spatially extended CR injection and Monte Carlo transport to provide an up-to-date and physically motivated model of the multi-messenger diffuse emission from galaxy clusters, the details are as follows.

\subsection{SLOW MHD simulation}\label{sec:mhdsim}
To constrain ICM, we explored the SLOW-MHD simulation \citep{dolag2023simulating, boss2024simulating}.
This simulation is designed to study the anomalies in the local Universe up to a distance $500\, h^{-1}\, \mathrm{Mpc}$.
The cosmological parameters in the simulation are: $\Omega_m=0.307, \, \Omega_\Lambda=0.693, \, H_0 = 67.77 \,\mathrm{km} \, s^{-1}\, \mathrm{Mpc}^{-1}, \, \sigma_8= 0.829$ based on Planck cosmology \citep{ade2016planck}.
The Initial conditions are derived based on peculiar velocities in the CosmicFLow catalog \citep{tully2023cosmicflows}.
This simulation is performed based on the GADGET code \citep{springel2005cosmological}.
AGN feedback and star-formation are not included in this simulation suite.
The galaxy clusters within the simulation volume generally reproduce the observed scaling relations between mass, temperature, X-ray luminosity, and the Sunyaev–Zeldovich signal \citep{singh2020cosmology, gupta2017sze}, enabling the cross-identification between observational signals and simulations.
Clusters such as Virgo, Norma, Perseus, Coma, Fornax, Hydra, and Centaurus are the key structures of the SLOW MHD simulation.
We focus on the Virgo, Perseus, and Coma clusters, as these systems have available high-energy $\gamma-$ray observations, with recent upper-limit estimates reported by LHAASO.
The predicted mass and the thermal energy of these structures by the SLOW simulation match well with observation-based estimates. 
The Coma cluster is one of the most massive galaxy clusters, has the virial mass of $\sim 1.4\times 10^{15}\, M_{\odot}$
and the thermal energy of $\sim 2\times 10^{64}$~erg. The Perseus cluster is somewhat less massive than Coma, with a total mass of  $\sim 7\times 10^{14}\, M_{\odot}$ and a thermal energy of $\sim 3\times 10^{63}$~erg. The Virgo cluster has a total mass of about  $\sim 5\times 10^{14}\, M_{\odot}$ and a thermal energy of 
 $\sim 2\times 10^{62}$~erg.
These values are estimated from the SLOW simulation, which are in good agreement with the Planck survey \citep{ade2011planck, ade2016planck}.
%
%

\begin{figure}[tbp]
\centering 
\includegraphics[width=.45\textwidth]{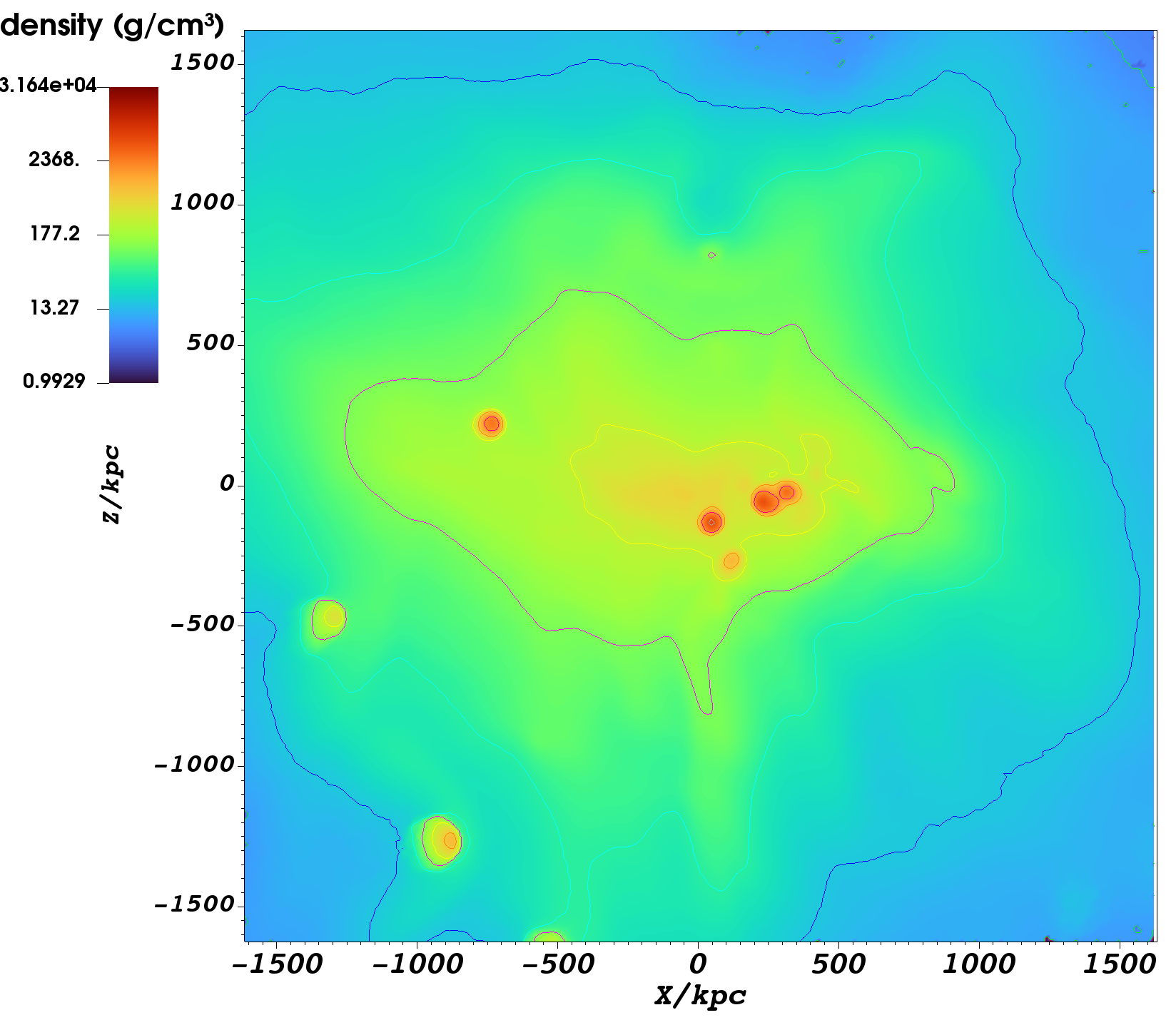}
\hfill
\includegraphics[width=.45\textwidth]{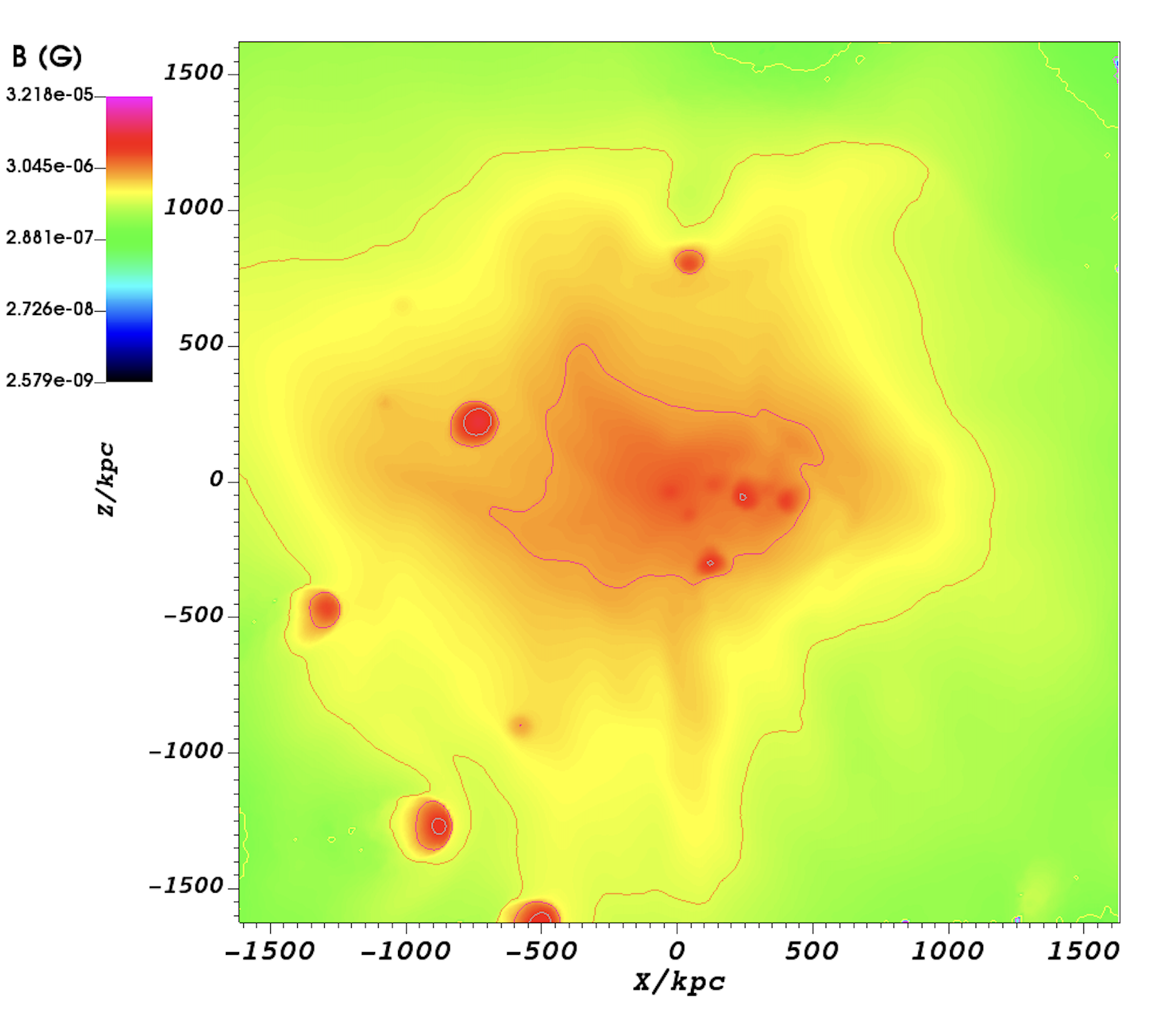}
\hfill
\includegraphics[width=.45\textwidth]{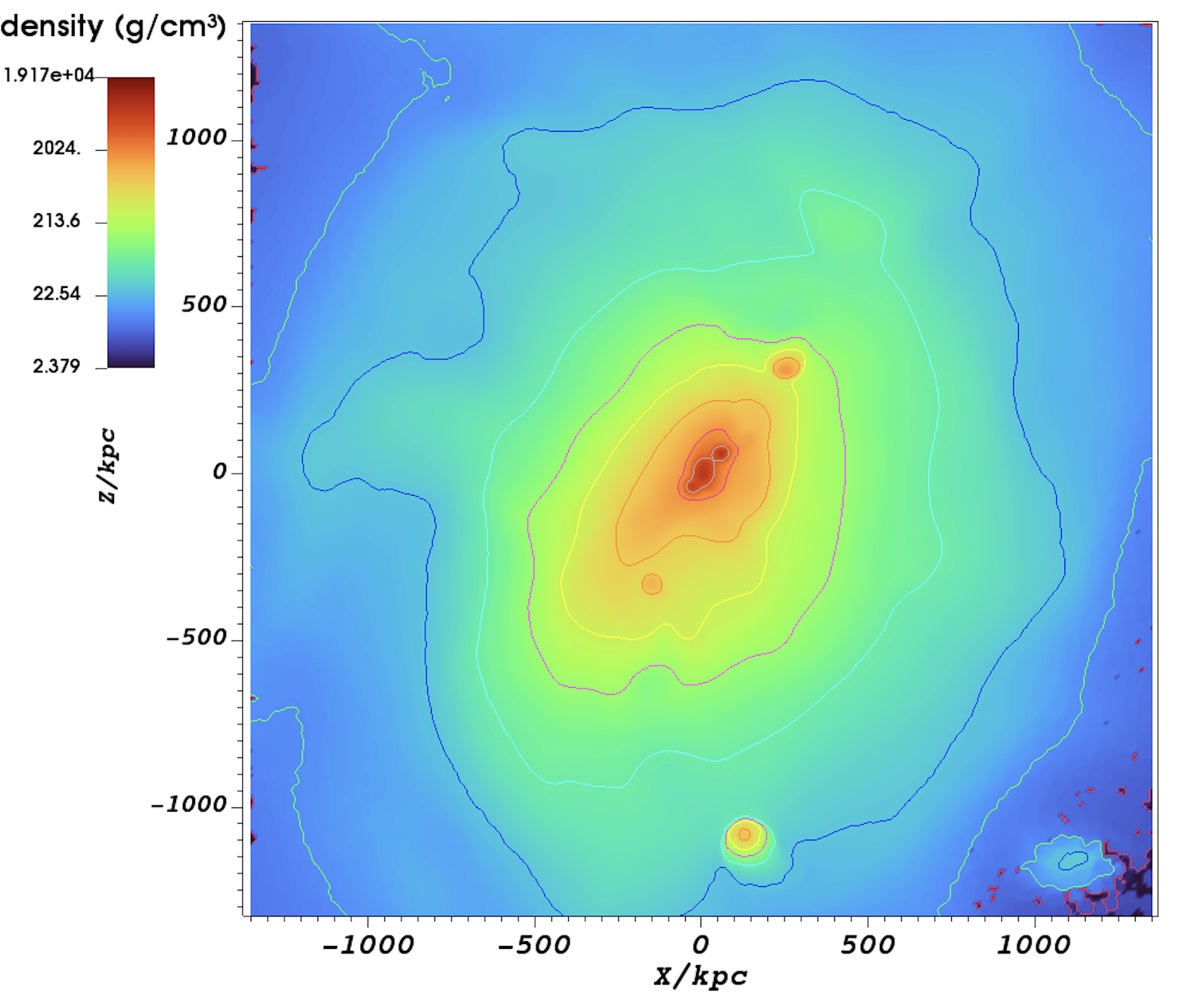}
\hfill
\includegraphics[width=.45\textwidth]{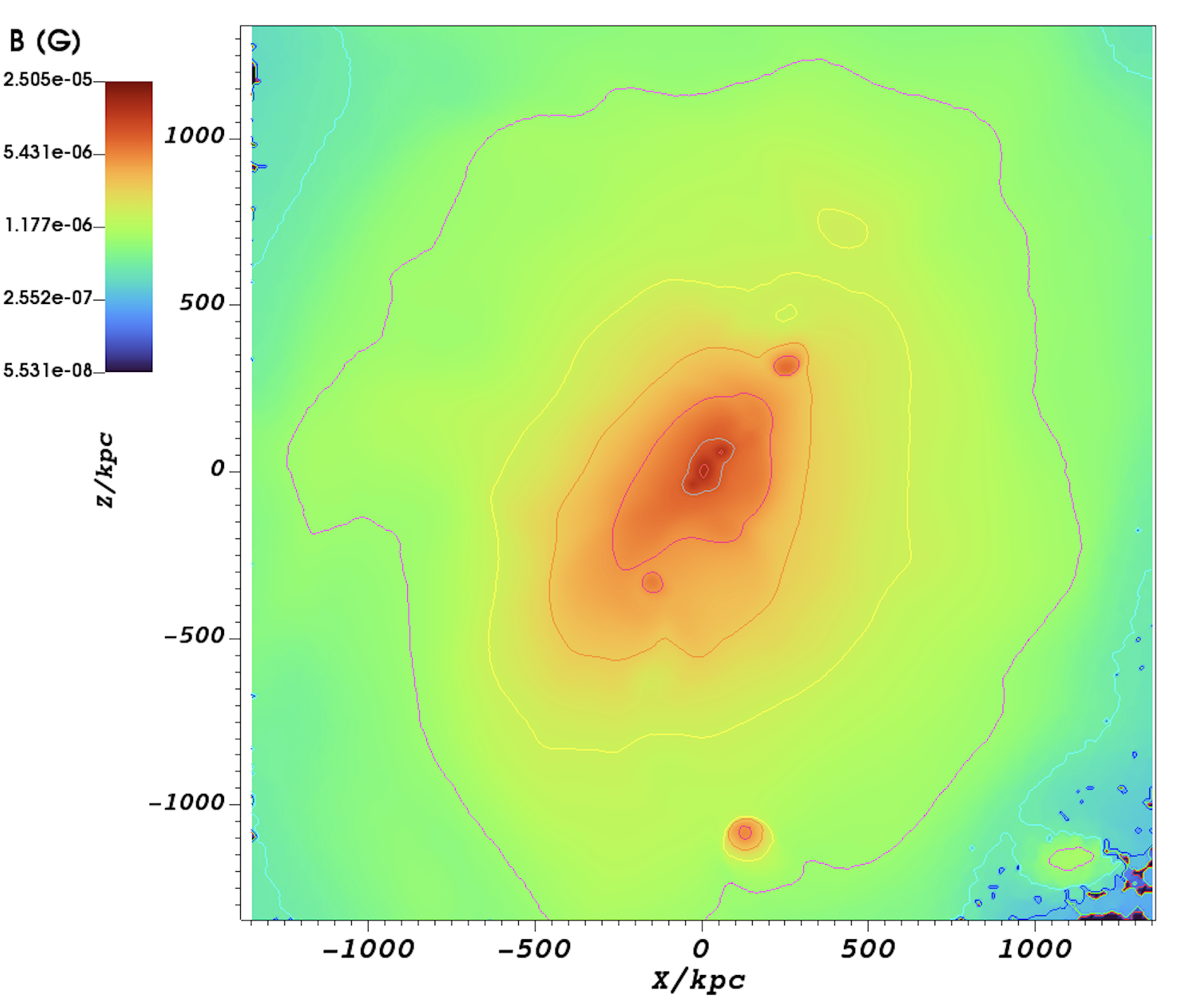}
\hfill
\hfill\includegraphics[width=.45\textwidth]{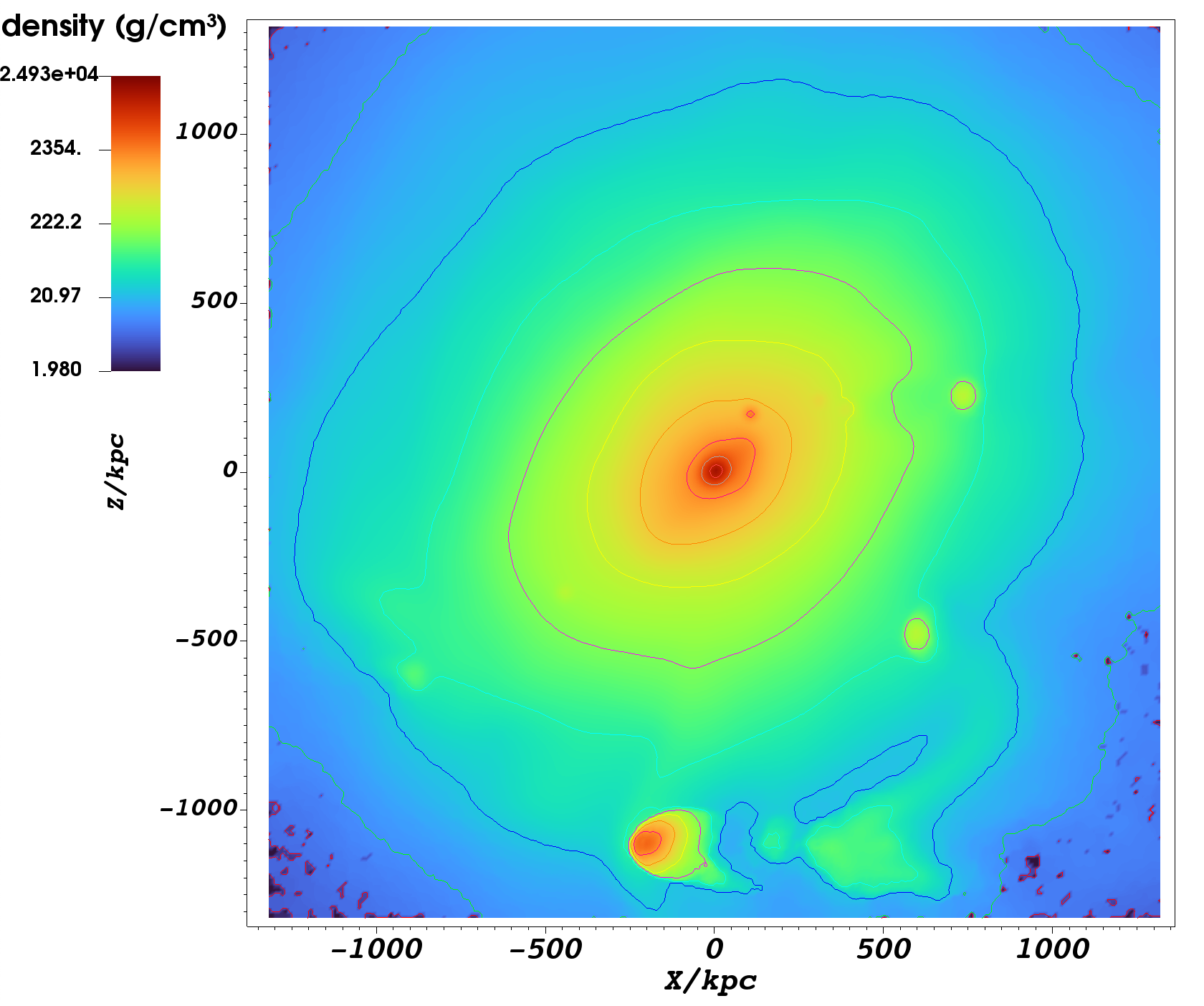}
\hfill
\includegraphics[width=.45\textwidth]{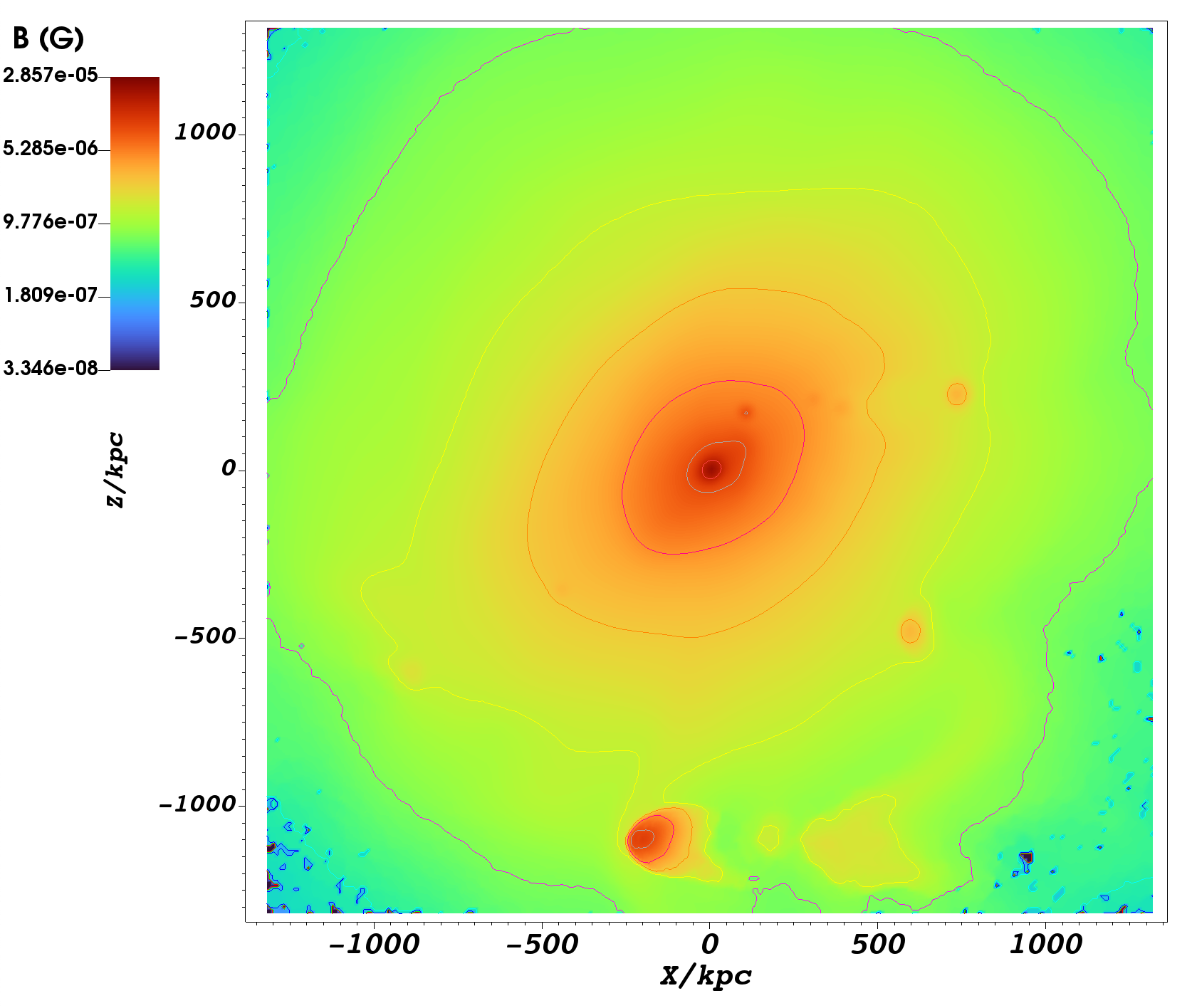}
\caption{\label{fig:densitymap-clusters} The left column (top to bottom) represents 2D (XZ) central slices (at y=0) of the gas density of the Coma, Perseus, and Virgo clusters, respectively. The right column represents the magnetic field configurations of the same clusters. These maps are obtained from the SLOW MHD simulation at redshift $z\simeq0$ \citep{dolag2023simulating}.}
\end{figure}

We obtained profiles of the temperature, magnetic field, and gas density of clusters directly from the MHD simulations.
Fig.\ref{fig:densitymap-clusters} represents two-dimensional (2D) central slices (XZ) for gas and magnetic field of the Coma, Perseus, and Virgo clusters, respectively. These 2D maps indicate the non-symmetric distributions of gas and magnetic field in clusters.
In particular, the gas density and magnetic-field maps of the Coma cluster (Fig. \ref{fig:densitymap-clusters}) show stronger asymmetries,
indicating that non-spherical morphologies are more pronounced in dynamically disturbed systems.

\subsection{CR Injection}\label{subsec:CRInject}

CRs are injected with a power-law energy distribution with spectral index $\alpha$ and exponential cut-off energy $E_{max}$, following the relation:
\begin{equation}
dN/dE\propto E^{-\alpha} exp(-E/E_\mathrm{max}).
\end{equation}
More importantly, the morphology of CR injection follows the gas density of galaxy clusters.
This approach provides a physically motivated prescription for CR injection that reflects the thermodynamic structure of the ICM.
The rationale behind this assumption is that CRs in clusters can be accelerated by AGN jets, merger and accretion shocks, and MHD turbulence. AGN activity and turbulence are often associated with dense central regions of clusters \citep{beckmann2022cosmic, beduzzi2024cosmological}, providing a natural connection between CR injection and the gas density. Since the gas density profile declines with radius, the injection rate is correspondingly higher in central regions, while remaining non-zero in the outskirts.
At the same time, merger shocks are frequently observed near radio relics at $\sim$Mpc scales \citep{vazza2012turbulence, vazza2015electron}, accretion shocks form around the virial radius \citep{zhu2021shock}, and turbulent pressure can increase toward cluster outskirts \citep{truong2024x}. Therefore, our injection model captures the overall relationship between CR production and the ICM without explicitly modeling the time-dependent evolution of individual shocks.

%
%

We show the CR injection in the simulated cluster using two complementary representations: a 2D energy–radius map illustrating the dependence on radius and energy, and a radial profile comparing a central injection with an extended injection following the volume-averaged gas density of a Perseus-like cluster as shown in Fig. \ref{fig:CRInjection} in the appendix.

%
%
%
%
%

The primary source of energy heating ICM to $10^8$~K is the merging of smaller structures (galaxies and subclusters) to larger ones, converting the gravitational potential energy to thermal and kinetic energy through shocks and turbulence \citep{norseth2025electron}.
%
A fraction of the energy released by these processes can be transferred to accelerate CRs. Determining the percentage of this energy that contribute to CR acceleration is one of the crucial points in this study. 
We leave it as a free parameter to be regulated by observations and MHD simulations.
The CR energy budget in galaxy clusters can be represented as the ratio $X_\mathrm{CR} = E_\mathrm{CR}/E_\mathrm{cluster}$, where $E_\mathrm{CR}$ is the integrated energy of CRs and $E_\mathrm{cluster}$ is the total thermal energy of the clusters. 
The $X_\mathrm{CR}$ depends on AGN activities and shocks inside the clusters. It also depends on the masses and redshifts of the clusters.
Following the $\gamma$-ray constraints reported by \citep{ackermann2014search}, the non-detection of spatially extended emission from nearby galaxy clusters implies an upper limit on $X_\mathrm{CR}$ at the level of a few percent. In this work, we adopt $X_{\rm CR} \simeq 0.01$ as a fiducial model parameter, which lies within these observational limits and is consistent with values commonly assumed in the previous studies \citep[e.g., ][]{fang2016high, fang2018linking}. 
%
Therefore, to convert the simulation results to physical units, we fixed $X_{CR}=0.01$ for all clusters considered in this work, including the Virgo, Perseus, and Coma.


\subsection{CR diffusion}
To investigate CR transport in a diffusive regime, CRPropa uses the Parker transport equation,
\begin{equation}\label{eq:SDE-parker}
\frac{\partial n}{\partial t} + \vec{u} . \nabla n = \nabla.(\hat{k} \nabla n) + \frac{1}{p^2}\frac{\partial}{\partial p} \big(p^2 \kappa_{pp} \frac{\partial n}{\partial p} \big) + \frac{1}{3}(\nabla.\vec{u})\frac{\partial n}{\partial \ln p} + S(\vec{x}, p, t),
\end{equation}
which is a simplified version of the Fokker-Planck equation.
Here $n$ is the particle density, $\vec{u}$ is the advection speed, $p$ is the absolute momentum. $\kappa_{pp}$ is the momentum diffusion coefficient that describes the re-acceleration, $\hat{\kappa}$ is the spatial diffusion tensor.
The mean free path (MFP) associated with spatial diffusion in Eq. \ref{eq:SDE-parker} is treated as an effective quantity describing CR scattering by magnetic turbulence in the ICM, within the standard Parker transport formalism. Given the large uncertainties in the turbulence spectrum, magnetic-field structure, and plasma properties of galaxy clusters, this scattering MFP is modeled phenomenologically, following common practice in cluster-scale CR transport studies (e.g.~\cite{vazza2017simulations, merten2017crpropa}). The particle density $n$ evolves under the combined effects of advection, spatial diffusion, momentum diffusion, and source injection. The spatial diffusion tensor and the momentum diffusion coefficient represent effective transport parameters describing CR scattering and stochastic re-acceleration by magnetic turbulence, and are implemented using stochastic differential equations \citep[see e.g.,][]{gardiner2009stochastic, merten2017crpropa}. These transport processes are defined on top of the background gas density and magnetic-field distributions described above, which are obtained from the SLOW simulation and treated as static snapshots at $z\simeq0$. The microphysical interaction of CRs with turbulent fluctuations is therefore not resolved explicitly, but its net effect on CR transport is captured through the effective diffusion coefficients.

In this framework, the source term $S(\vec{x}, p, t)$ represents the local phase-space injection rate of CRs and provides a continuous description of the injection process defined above (Sec. \ref{subsec:CRInject}). It encodes the spatial distribution of CRs, the spectral shape, and the maximum energy, and the normalization of the injected particles. It also determines how CRs are introduced into the transport equation as a function of position, momentum, and time. In CRPropa, this continuous source term is realized through an ensemble of pseudo-particles whose properties are sampled from the prescribed injection distributions, which ensure a consistent numerical realization of the source term and its coupling to propagation, diffusion, and re-acceleration processes \citep[see e.g.,][]{merten2017crpropa, batista2022crpropa}.

For a $1\,\mathrm{PeV}$ proton in a  intracluster magnetic field of 
$B \sim 1\,\mu\mathrm{G}$, the Larmor radius is 
$r_{\rm L} \approx 1\,\mathrm{pc}$, which is much smaller than the magnetic 
coherence length in galaxy clusters ($l_c \sim 10$-$50\,\rm{kpc}$). 
Transport therefore proceeds in the resonant scattering regime, where 
pitch-angle scattering is governed by magnetic turbulence at spatial scales 
comparable to $r_{\rm L}$. 
Under realistic ICM conditions, where the magnetic turbulence level is 
$\delta B/B < 1$, scattering is weaker and the MFP is 
correspondingly larger, $\lambda \sim 10$--$100\,\mathrm{pc}$ in cluster cores, 
increasing toward the outskirts as the magnetic field strength declines. 
In our framework, the effective scattering scale is parameterized by the adopted 
diffusion coefficient, chosen to be consistent with the magnetic field strengths 
and turbulence levels expected from the underlying MHD simulations.
The diffusion coefficient is not inferred from a direct analysis of turbulence in the MHD simulations; instead, it is implemented following the standard prescription in CRPropa. In this numerical setup, the parameter $\epsilon$ is defined as the ratio between the perpendicular and parallel diffusion coefficients, $\epsilon = \kappa_{\perp}/\kappa_{\parallel}$; we adopt $\epsilon = 0.1$, consistent with typical expectations for the ICM \citep{shalchi2022ratio}.
At sufficiently high energies ($>$ PeV), CR transport approaches a semi-diffusive regime, and the results become less sensitive to the specific choice of diffusion parameters \citep{condorelli2023impact}.

Under typical ICM conditions and for particle energies $\gtrsim \rm{PeV}$, the dynamical impact of stochastic (second-order Fermi) re-acceleration is limited.
For $\mu$G-level magnetic 
fields and Alfv\'en speeds $v_A \sim 100\,\mathrm{km\,s^{-1}}$, the 
characteristic acceleration timescale 
$t_{\rm acc} \sim D/v_A^2$ becomes increasingly long toward high energies, 
reaching $t_{\rm acc} \gtrsim 10^8$--$10^9\,\mathrm{yr}$ at PeV and above, 
comparable to or exceeding cluster dynamical timescales. Consequently, 
Stochastic re-acceleration is inefficient, particularly at the highest energies considered.
Diffusive shock acceleration (DSA) at merger or accretion shocks 
($u_s \sim 10^3\,\mathrm{km\,s^{-1}}$) can in principle contribute to 
particle energization; however, cluster shocks are typically weak 
(Mach number $\mathcal{M} \lesssim 2$--$4$), limiting their efficiency. 
Moreover, our simulations adopt a static ICM background derived from MHD simulations and 
therefore do not model time-dependent shock evolution or explicit shock 
injection.

Depending on the Larmor radius $(r_L \approx 1.08\, E_{15}/B_{\mu \mathrm{G}}\, \mathrm{pc})$, CR trajectories can be diffusive or semi-diffusive. For a CR of energy $\sim 10^{17}\, \mathrm{eV}$ and $B= 1\mu$~G, the Larmor radius is $\sim 0.1$~kpc, which is much smaller than the size ($\sim 1$~Mpc) of a galaxy cluster. So, we are in a diffusive regime. 
In principle, CRs with energy $\lesssim 10^{17}$~eV could be confined in clusters for a time comparable to the Hubble time ($t_H \approx 13$~Gyr) and CR with higher energies have more chances of escape.
Using the Hubble time as a reference, the maximum trajectory length of a CR is $\sim 10^{3}$~Mpc, and the confinement time can be calculated as $t_\mathrm{con}\approx 1000\,\mathrm{Mpc}/c\approx t_\mathrm{H}$. This trajectory represents the total distance a CR travels inside a cluster before it reaches the observer.
The escaping CR flux depends on the mass and magnetic field configuration of the clusters;
more massive clusters possess stronger magnetic fields that confine high-energy CRs efficiently, thereby enhancing the production of secondary particles.

\subsubsection{CR propagation in ICM}\label{subsec:CRPropagation}
CR propagation in ICM is studied using the Monte Carlo code CRPropa \citep{batista2022crpropa}. 
The background magnetic field, gas density, and temperature distribution for clusters of different masses and redshifts are obtained from the SLOW simulation \citep{dolag2023simulating}.
During the propagation of CRs inside and outside of clusters, we considered all the relevant CR interactions, including photopion production, photodisintegration, Bethe-Heitler pair production, and pp-interactions. The electromagnetic cascade processes, including pair production and inverse Compton scattering (ICS), are also considered.
In our simulations, only CR protons are injected, while electrons and positrons are produced self-consistently as secondary products of hadronic interactions. ICS is computed during the propagation of these secondary leptons.
The dominant seed photon field is the cosmic microwave background (CMB), which provides an isotropic and spatially uniform radiation field throughout the ICM, while the EBL is included as a sub-dominant component. Under typical cluster conditions, scattering on the CMB dominates the resulting $\gamma-$ray emission, with the EBL contributing mainly at the highest electron energies.
For example, TeV-scale secondary electrons upscatter CMB photons into the GeV–TeV $\gamma-$ray band, which is the main energy range considered in this work.
Our simulations also account for the synchrotron losses and adiabatic losses due to the expansion of the Universe.
The photon fields considered in this work are the CMB, extragalactic background light (EBL), and radio background \citep{protheroe1996new}.
For EBL, we used the Gilmore model \citep{gilmore2012semi}, other models like \cite{dominguez2011extragalactic} also provide similar attenuation effects \citep{hussain2023diffuse}.
The flux of CRs, neutrinos, and $\gamma-$rays collected at the edge of clusters and then propagated through the IGM depends on the distance of each cluster in the SLOW simulation.

In our simulations, we assume that CRs consist solely of protons, which is a reasonable approximation because the contribution of heavy ions to $\gamma-$rays and neutrinos would be subdominant \cite[see e.g., ][]{kotera2009propagation}.
The intergalactic magnetic field (IGMF) is not considered during the propagation of $\gamma-$rays outside clusters, as it is not presumed to significantly affect the $\gamma-$ray flux above $10$~GeV \citep{alves2021gamma}.

The most important channel to produce neutrinos and $\gamma-$rays is pp-interaction. However, above $10^{17}$~eV the photopion production becomes dominant \cite[for details see][]{hussain2021high}.
The pp-interactions are modeled explicitly using a dedicated module outside the core CRPropa framework.
It models the interaction of high-energy CRs with hydrogen-rich gas in astrophysical environments.
The pp-interaction and the corresponding spectra of secondary particles are parameterized following \cite{kelner2006energy, kafexhiu2016parametrization}, 
see appendix \ref{appen:ppInter} for details.
%
%


\section{Results}\label{sec:results}
\subsection{$\gamma-$ray and Neutrino Fluxes from Individual Clusters}
The $\gamma$-ray (and neutrino) flux associated with individual galaxy clusters is computed directly from Monte Carlo simulations performed with CRPropa code. 
The simulations use the SLOW-MHD outputs as a background for the three-dimensional gas density and magnetic field distributions. 
CRs are injected following the local gas density, and their propagation is self-consistently tracked both inside and outside the cluster. 
All relevant processes are included, such as hadronic interactions, electromagnetic cascading, and propagation through the intergalactic medium (as described in Sec. \ref{sec:method}).

\begin{figure}[tbp]
\centering 
\includegraphics[width=.45\textwidth]{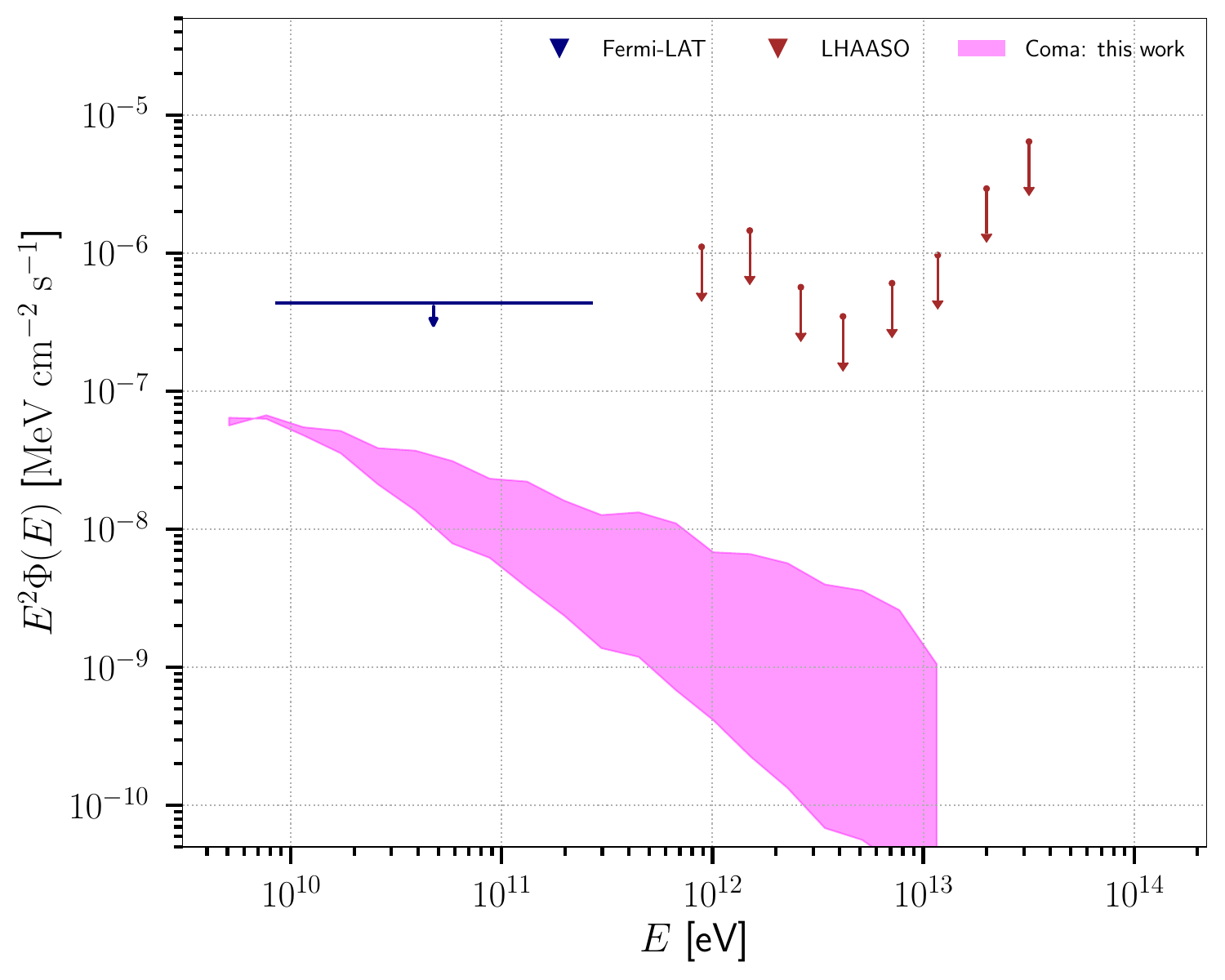}
\hfill
\includegraphics[width=0.45\textwidth]{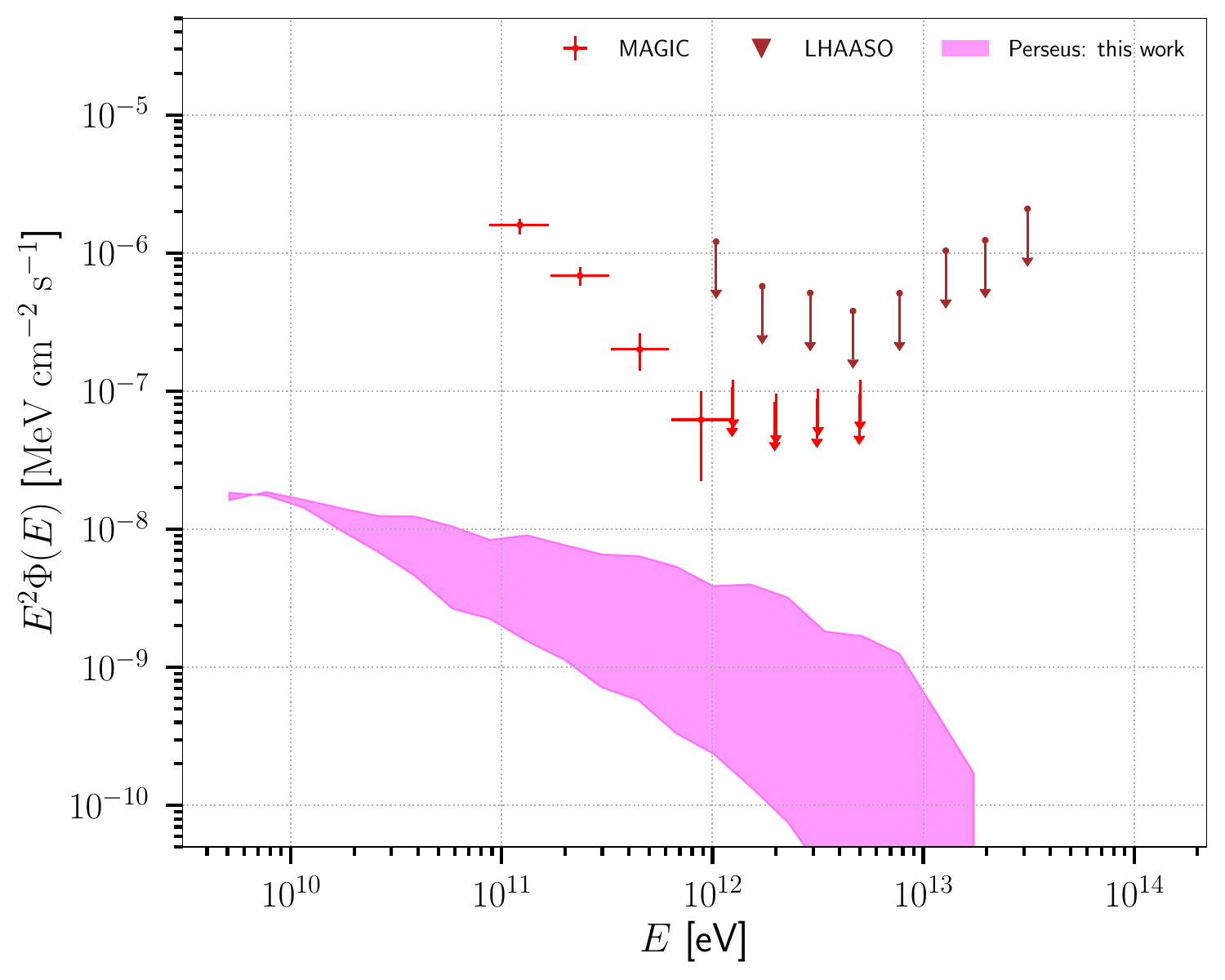}
\hfill
\includegraphics[width=0.45\textwidth]{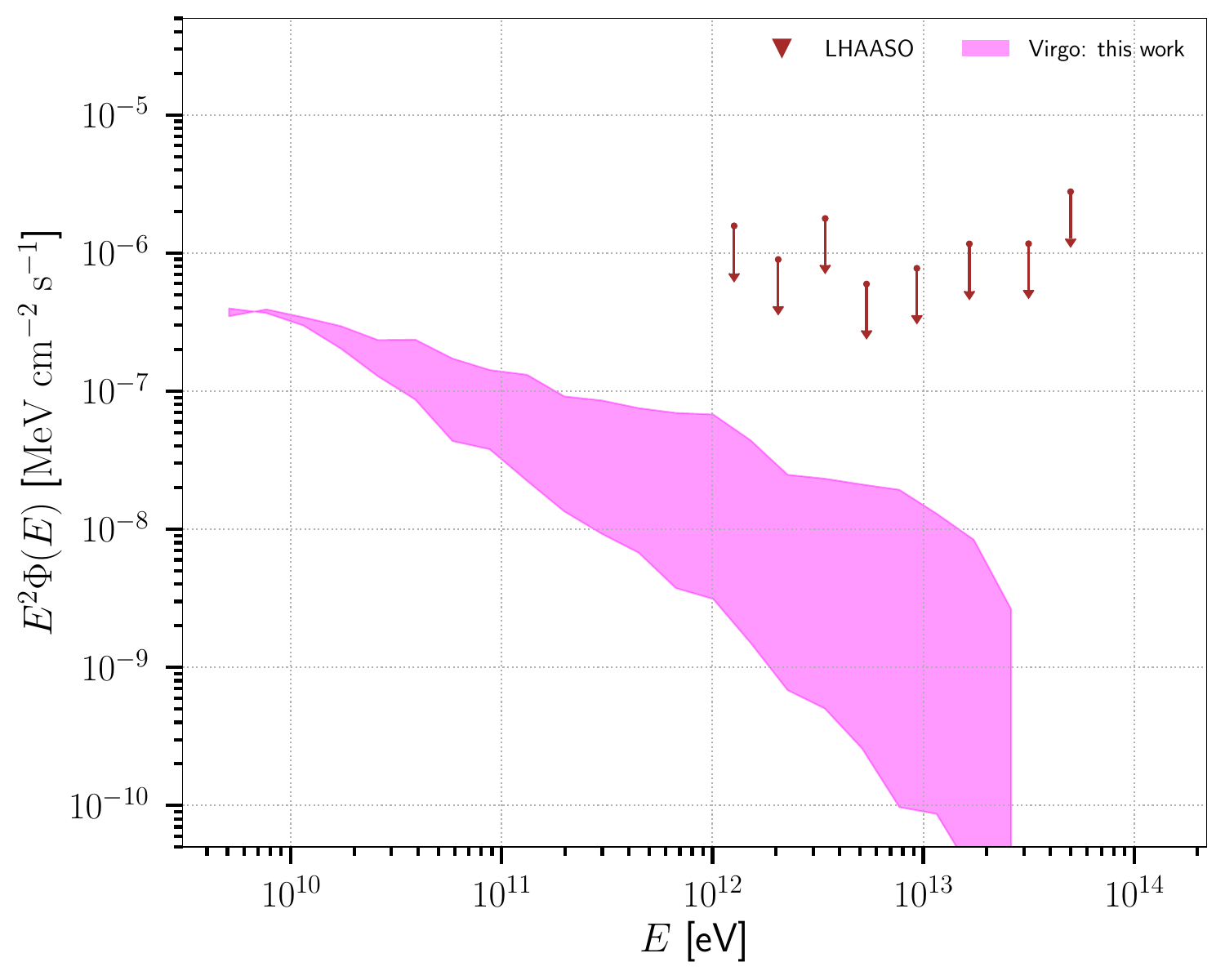}
\caption{\label{fig:gammaClusters} The $\gamma-$ray spectra in our simulations, integrated over the entire volume of the Coma, Perseus, and Virgo clusters. The magenta band represents spectra for parameters $\alpha = 2.0-2.5\, \&\, E_\mathrm{max}= 10^{16}- 10^{17}\, \mathrm{eV}$. We also depicted upper limits of LHAASO for these clusters \citep{cao2025constraining}. Blue error bar in the top left panel represents the upper limits of Fermi-LAT for the Coma cluster above energy $10$ GeV \citep{xi2018detection}. The top right panel also shows the $\gamma-$ray flux observed by MAGIC from the central region (NGC$1275$) of the Perseus cluster \citep{ahnen2016deep}.}
\end{figure}

Fig. \ref{fig:gammaClusters} shows the $\gamma-$ray flux for Virgo, Perseus, and Coma clusters, it is presented for parameters $\alpha = 2.0-2.5\, \&\, E_\mathrm{max}= 10^{16}- 10^{17}\, \mathrm{eV}$ and $X_{CR}\approx 0.01$.
The $\gamma-$ray flux obtained from our simulations for the Coma, Perseus, and Virgo clusters is lower by an order of magnitude than the recent upper limits reported by the LHAASO Collaboration \citep{cao2025constraining}. 
The LHAASO upper-limits in the energy range $(1-100)$~TeV  are obtained for individual clusters by modeling the emission as a disk of radius $R_{500}$ and without explicitly separating diffuse intracluster emission from central or embedded sources \citep{cao2025constraining}. Consequently, these limits are less constraining for diffuse hadronic emission from the ICM and lie well above our model predictions.
In contrast, our study focuses exclusively on the diffuse component, the emission produced by large-scale CR interactions within the ICM.

The MAGIC collaboration had studied the region of the Perseus cluster and observed the $\gamma-$ray emission in the TeV energy range from the central source NGC1275 \citep{ahnen2016deep}. 
However, they did not detect any diffuse emission signal, providing constraints on the CR population (CR to thermal pressure) in the Perseus cluster. 
Also, the SHALON experiment observed TeV $\gamma-$rays from the NGC1275 \citep{sinitsyna2014emission}. The observations by MAGIC and SHALON are compatible around the TeV energy, but exceed our predicted fluxes.
The MAGIC and LHAASO limits for the Perseus cluster overlap at TeV energy, but the predicted LHAASO limits are slightly higher.
It would be challenging to significantly improve the current constraints regarding diffuse $\gamma-$ray emissions from Perseus-like clusters with the existing generation of Cherenkov telescopes.
However, the upcoming CTA is expected to provide more stringent limits.
Our results can serve as a reference for future observations by instruments such as CTA and LHAASO, etc.

%
For the Coma cluster, considering an extended spatial structure, the upper limits estimated by \cite{xi2018detection}, in the energy bin $\sim(10-300)$~GeV using the Fermi-LAT data, exceed our results roughly by an order of magnitude.

\begin{figure}
\centering
\includegraphics[width=0.5\linewidth]{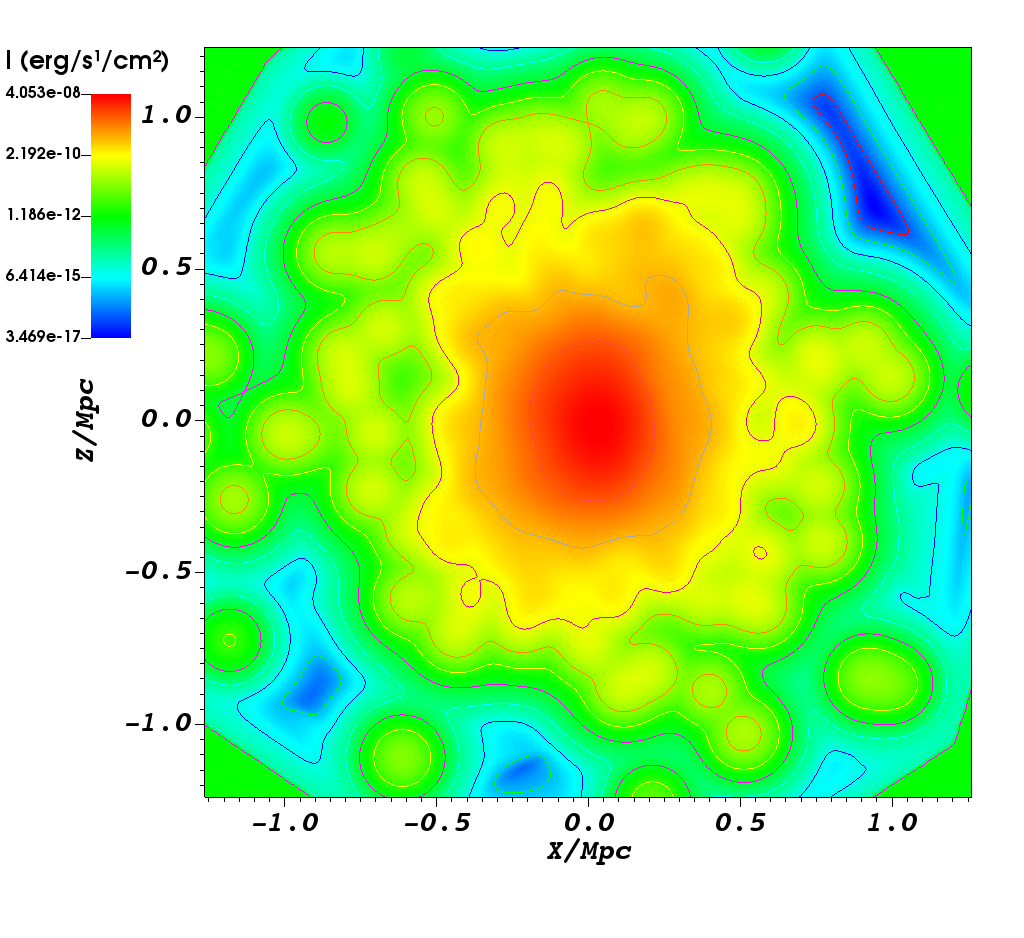}
\caption{Projected 2D $\gamma-$ray intensity map of a Perseus-like cluster obtained directly from our Monte-Carlo simulations.
}\label{fig:intensityPersesu}
\end{figure}

We compute the projected $\gamma$-ray intensity map directly from the Monte Carlo simulations for a Perseus-like cluster, shown in Fig. \ref{fig:intensityPersesu}. 
The projected $\gamma-$ray intensity map for the Perseus-like cluster is broadly centrally peaked but exhibits noticeable small-scale asymmetries and clumpy features compared to a simple spherically symmetric model.
We compute the projected radius enclosing $95\%$ of the total $\gamma$-ray intensity, for $\gamma$-rays with energies $\geq 10\,\mathrm{GeV}$, we find $R_{95} \simeq 1.0\,\mathrm{Mpc}$. 


\begin{figure}[tbp]
\centering 
\includegraphics[width=.5\textwidth]{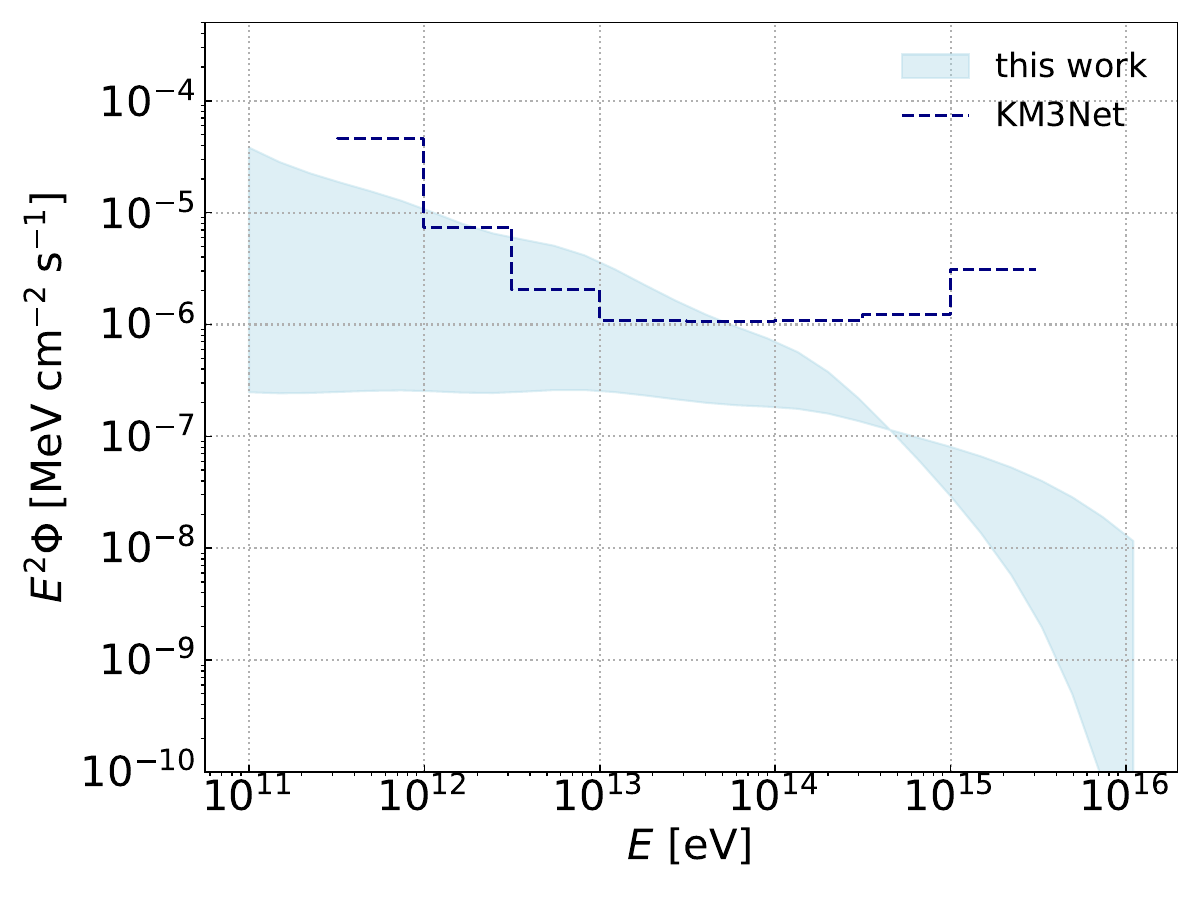}
\caption{\label{fig:neutrino_15Mpc}All-flavor neutrino flux for Virgo cluster, for same parameters as in Fig. \ref{fig:gammaClusters} ($\alpha = 2.0-2.5\, \&\, E_\mathrm{max}= 10^{16}- 10^{17}\, \mathrm{eV}$).  
We also show the 90\% confidence level differential sensitivity of KM3NeT, calculated for an all-flavor neutrino flux \citep{aiello2024differential}.
}
\end{figure}

In Fig. \ref{fig:neutrino_15Mpc}, we present the all-flavor neutrino flux for the Virgo cluster
predicted by our simulation.
We compared our results with the differential sensitivity of $90\, \%$ confidence level of the Cubic Kilometre Neutrino Telescope (KM3NeT) \citep{aiello2024differential}). 
The KM3NeT/ARCA differential sensitivities are calculated using Monte Carlo simulations that include all neutrino flavors combined.
Our simulated neutrino flux for the Virgo cluster is comparable to the expected sensitivity of the KM3NeT/ARCA detector, indicating that the Virgo cluster is a particularly promising target for future observational campaigns targeting diffuse neutrino emission. 
Consequently, this work provides a theoretical framework that motivates dedicated searches for diffuse neutrinos from the Virgo cluster with experiments such as KM3NeT and IceCube.
%
%

\subsection{Multi-messenger Emission from Local Galaxy Clusters}
We calculated the total $\gamma-$ray and all-flavor neutrino fluxes from galaxy clusters of mass range $\gtrsim 5\times10^{13} M_{\odot}$ within $500\, h^{-1}$~Mpc comoving distance.
The integrated neutrino and $\gamma-$ray fluxes of clusters within a $500\, h^{-1}$~Mpc distance and mass range $5\times10^{13} \lesssim M/M_{\odot}\lesssim 2\times 10^{15}$ are calculated as:
\begin{equation}\label{eq:flucCal}
E_\mathrm{obs}^{2} \Phi(E_\mathrm{obs}) =  \int\limits_{z_\mathrm{min}}^{z_\mathrm{max}} dz \int\limits_{M_\mathrm{min}}^{M_\mathrm{max}} dM \frac{dN}{dM} E^{2}  \frac{d\dot{N}( E/(1+z), M , z)}{dE}  \times
g(E_\mathrm{obs}, E, z)
\left(\frac{1}{4\pi d_\mathrm{com}^2 (z)}\right) 
\end{equation}
where $dN/dM$ is the number of clusters per mass interval calculated from the SLOW simulation \citep{hernandez2024simulating}, $g(E_\mathrm{obs}, E, z)$
accounts for the interactions of gamma rays in the ICM and the IGM, the quantity $E^2 \; d\dot{N}/dE$ denotes the neutrino or $\gamma-$ray power spectrum obtained from the Monte-Carlo simulations, and $d_\mathrm{com}$ is the comoving distance.
Throughout this work, we adopt the same cosmological parameters as assumed in the SLOW simulations see Sec. \ref{sec:mhdsim}.
The integration limits adopted in Eq. \ref{eq:flucCal} are 
$z_{\rm min} = 0.0036$ (corresponding to the redshift of the Virgo cluster) and $z_{\rm max} = 0.12$ for the redshift range, 
and $M_{\rm min} = 5 \times 10^{13}\,M_\odot$ and 
$M_{\rm max} = 2 \times 10^{15}\,M_\odot$ for the cluster mass interval. 
These values are chosen to match the parameter space covered by the 
underlying MHD SLOW simulation. 
The SLOW 
simulation reproduces key observational properties of galaxy clusters 
and thus provides a physically motivated and observationally consistent 
framework for defining the integration bounds.

Eq.~\ref{eq:flucCal} provides a convenient parametrization of the cluster $\gamma-$ray and neutrino fluxes as a function of mass and redshift. 
This parametrization is not imposed \emph{a priori}, but summarizes the outcome of Monte Carlo simulations performed with CRPropa, which model CR transport and interactions in realistic cluster environments provided by the SLOW simulation. 
The dependence on mass and redshift emerges naturally from the simulated physical processes, including the cluster gas content, magnetic-field–dependent CR confinement, interaction rates, and cosmological propagation effects (see Sec. \ref{subsec:CRPropagation} for details).
To illustrate how the predicted $\gamma$-ray and neutrino fluxes depend on cluster properties, we perform Monte Carlo simulations for a representative set of clusters using physical conditions derived from the SLOW simulations. The resulting trends with cluster mass and redshift are presented in Appendix~\ref{append:mass-Redshift-depend}. In particular, lower-mass clusters, being more numerous, contribute efficiently to the $\sim10$~GeV $\gamma$-ray emission, whereas the high-energy ($\gtrsim 10$~TeV) neutrino flux remains largely dominated by massive systems due to their greater ability to confine the highest-energy CRs.

To compute the diffuse background, we include all clusters with mass range $\gtrsim 5\times 10^{13}\,M_\odot$ from the SLOW simulation \citep{hernandez2024simulating} within the comoving distance $500\, h^{-1}$~Mpc. For massive nearby clusters ($M_{500} > 7\times10^{14}\,M_\odot$), such as Virgo, Coma, and Perseus, we sum their fluxes individually. For the broader cluster population within $500$~Mpc ($z \lesssim 0.1$), we retain the integral over redshift and mass (Eq.~\ref{eq:flucCal}), using comoving distances provided by the SLOW simulation. Because the simulation volume contains sufficiently large number of clusters above $5\times 10^{13}\,M_\odot$, the total flux is well sampled and shot noise is negligible. 
The diffuse flux is then obtained as the sum of fluxes from individual nearby massive clusters plus the integrated contribution from the rest of the population, consistently capturing both discrete and collective contributions.

\begin{figure}[tbp]
\centering 
\includegraphics[width=.8\textwidth]{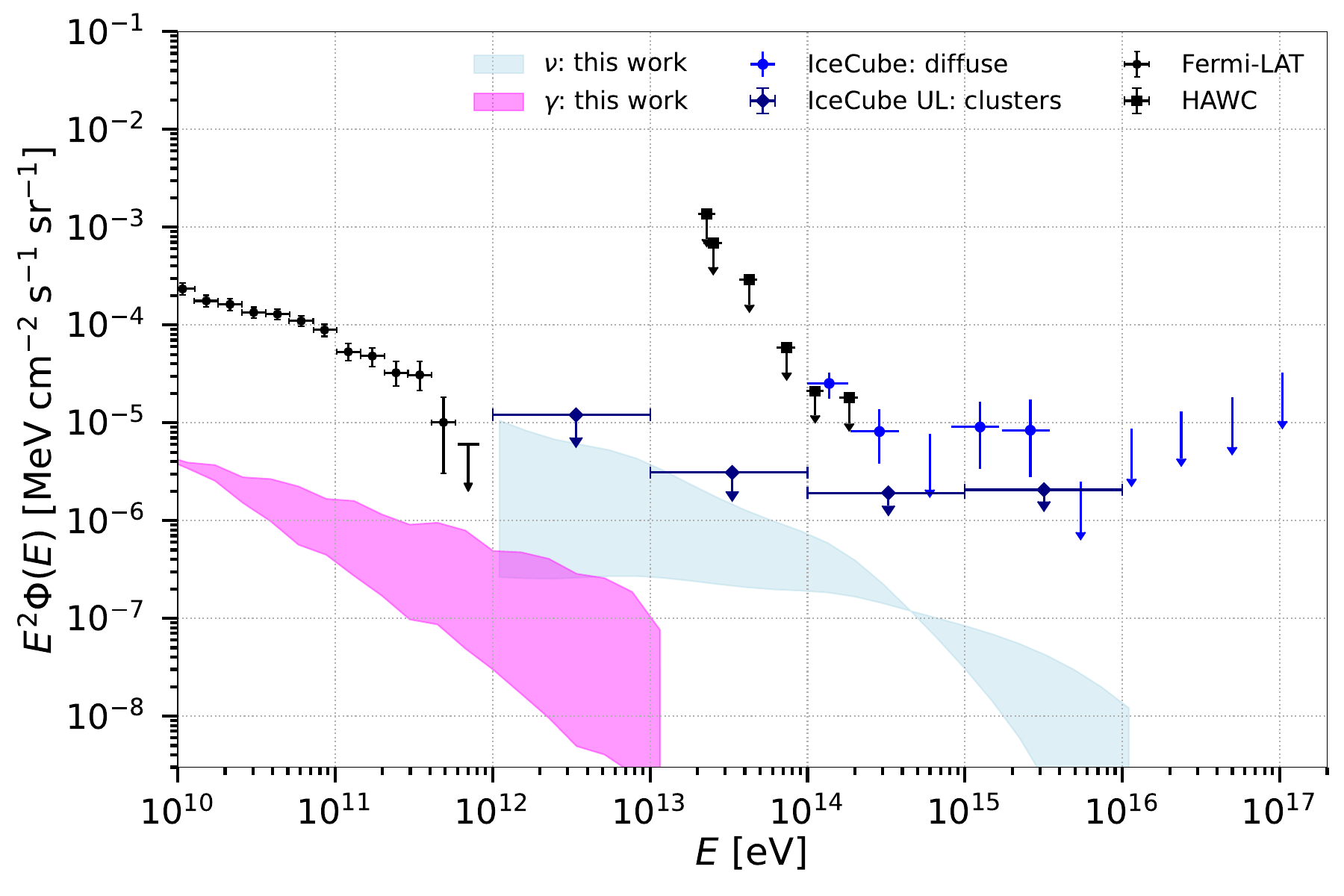}
\caption{\label{fig:totalGammaNeu}The magenta ($\gamma-$rays) and blue (neutrinos) bands represent spectra for parameters $\alpha = 2.0-2.5\, \&\, E_\mathrm{max}= 10^{16}- 10^{17}\, \mathrm{eV}$. These bands represent the integrated fluxes of $\gamma-$rays and  neutrinos (\SH{all-flavor}) computed from our simulations for all clusters of masses $\gtrsim 5\times 10^{13}\, M_{\odot}$ within a distance $\sim 500$~Mpc. The diffuse $\gamma-$ray flux of Fermi-LAT \citep{ackermann2015spectrum} and HAWC upper limits \citep{HAWC2022limits} are depicted as well as the diffuse neutrino flux \citep{aartsen2015evidence, aartsen2015searches} and upper limits \citep{abbasi2022searching} predicted by IceCube collaboration.   }
\end{figure}

Fig. \ref{fig:totalGammaNeu} shows the integrated flux of $\gamma-$ray and neutrino from clusters of mass range $> 5\times 10^{13} M_{\odot}$ within a comoving distance $500\, h^{-1}$~Mpc.  
The number density of clusters in this mass and distance range is $\sim 10^{-5}\,\mathrm{Mpc}^{-3}$, as estimated from the SLOW simulation, which is consistent with previous works \citep[e.g., ][]{tinker2010large,bocquet2016halo}.
Our $\gamma-$ray flux is plotted for the parameters $\alpha = 2.0-2.5\, \&\, E_\mathrm{max}= 10^{16}- 10^{17}\, \mathrm{eV}$ and $X_{CR}\approx 0.01$.
The diffuse $\gamma-$ray flux above $\sim 100$~GeV reported by Fermi-LAT \citep{FermiDGRB2015spectrum} is significantly higher than our results.
By including the contribution of relatively high redshift ($z>0.1$) clusters, our $\gamma-$ray flux would be comparable to the Fermi-LAT data. 
Below $\sim 100$~GeV, the $\gamma-$ray contribution of individual sources such as blazars \citep{Raniere2022isotropic}, misaligned AGN \citep{di2013diffuse}, and starburst galaxies \citep{roth2021diffuse} is dominant over clusters \citep{hussain2023diffuse}.

The total neutrino flux obtained from our simulation, from the same population of clusters, is well below the diffuse flux observed by IceCube \citep{aartsen2015evidence, aartsen2015searches}, but is comparable to the existing IceCube upper limits for clusters \citep{abbasi2022searching}. 
For comparison with IceCube muon-neutrino constraints, we assume standard flavor equipartition at Earth ($1:1:1$) and convert the all-flavor prediction to the corresponding muon-neutrino flux.
%
%
Note that we are not fitting the IceCube data, rather comparing our flux calculated from the entire population of clusters with masses $> 5\times 10^{13}\, M_{\odot}$ within a $500\, h^{-1}$~Mpc.

\section{Discussion}\label{sec:discussion}
In this work, we focus on hadronic $\gamma$-ray production in galaxy clusters. CR protons interact with ICM primarily through pp-collisions, producing neutral pions that decay into gamma rays, as well as charged pions that generate secondary electrons and positrons. These secondary leptons can account for the diffuse synchrotron emission observed in radio halos \citep{kushnir2024coma}, thereby linking radio observations to hadronic processes.
Although photopion interactions become relevant at proton energies $\gtrsim 10^{17}\, \rm eV$, contributing at the highest energies considered here, the $\gamma-$ray emission in the energy range of interest remains primarily governed by pp-interactions.

 
We modeled the production of hadronic $\gamma-$rays and neutrinos from three massive clusters, Virgo, Perseus, and Coma. 
Massive clusters ($\sim 7\times10^{14}\, M_{\odot}$) can confine CRs with energies up to $\sim 10^{17}$ eV. During confinement, the CRs interact with the gas and photon fields of the ICM, producing secondary particles.
The most important channel to produce $\gamma-$rays and neutrinos is pp-interactions, while above energy $>10^{17}$~eV, photopion production on the CMB and EBL becomes the dominant process. 
The slope of $\gamma-$ray and neutrino fluxes has a strong dependence on the CR injection spectrum, as expected.
We assume that CRs are accelerated in clusters through accretion or merger shocks; these phenomena suggest soft spectral indices ($\alpha\gtrsim 2.0$).
In this work, the parameters such as cut-off energies $E_\mathrm{max}=10^{16}-10^{17}$~eV, the spectral indices $\alpha=2.0-2.5$, and the ratio $X_\mathrm{CR}\approx0.01$ are consistent with previous studies \citep{pinzke2010simulating, FermiCluster2014search, fang2018linking, hussain2023diffuse, boss2025simulating}. 
These parameters are also consistent with previous studies for the Coma cluster \citep[see e.g., ][]{brunetti2014cosmic, brunetti2020second, nishiwaki2021particle}.
%
%

In essence, our results refer to the diffuse emissions from galaxy clusters as they are produced by large-scale CR interactions in the ICM, whereas the observations and upper limits reported by MAGIC and LHAASO are more consistent with point-source emissions originating from the sources in the center of clusters. 
Although the origin of the $\gamma-$ray and neutrino fluxes predicted in our work is diffusive, but the embedded sources in clusters such as AGN or starburst galaxies can generate similar spatial and spectral signatures.
In addition, key parameters such as CR energy, spatial distribution, injection spectra, compositions, and confinement time in the ICM remain poorly constrained.
Therefore, separating diffuse emission from the point source is very complex, especially in galaxy clusters.

The LHAASO collaboration recently reported the detection of $\gamma-$ray emission up to $\sim 20$~TeV from M87 \citep{cao2024detection}, the central galaxy of the Virgo cluster. They also estimated upper limits for the diffuse emission from the Virgo cluster \citep{cao2025constraining}. 
The measured flux from the point source M87 is comparable to the upper limits estimated by LHAASO for the Virgo cluster.
This similarity makes it particularly challenging to disentangle diffuse cluster emission from that associated with the central point source.
However, our simulated diffuse $\gamma-$ray flux arising from large-scale CR interactions in the Virgo cluster is marginally below the flux level reported by LHAASO. 
\cite{adam2021gamma} reported a $\gamma-$ray excess toward the Coma cluster, while \citet{harale2025excess} reported a similar excess toward Abell~$119$, based on template-based analyses of Fermi-LAT data. These results rely on specific assumptions about the spatial and spectral distribution of CR and remain sensitive to background modeling and the treatment of nearby point sources. Our predicted $\gamma-$ray flux for the Coma cluster at $\sim 10\,\mathrm{GeV}$ lies below $10^{-7}\,\mathrm{MeV\,cm^{-2}\,s^{-1}}$, which is comparable to the energy-binned Fermi-LAT upper limits \citep{ackermann2016search} and to the flux levels inferred in these recent studies. 
The slightly lower flux predicted in our work arises naturally from our CRPropa simulations, employing a spatially extended CR injection model that follows the gas density of clusters, leading to a more diffuse hadronic $\gamma-$ray signal.
Overall, our results remain consistent with current observational constraints and with the tentative nature of existing claims of diffuse $\gamma-$ray emission from galaxy clusters.

In \cite{nishiwaki2021particle}, they estimated the contribution of Coma-like clusters to all sky neutrinos and diffuse $\gamma-$rays using both hadronic and leptonic scenarios, and our results roughly match their predictions. 
Recently, \cite{boss2025simulating} estimated diffuse $\gamma-$ray emission from local galaxy clusters using a MHD simulation with an on-the-fly spectral CR model. Their results are largely compatible with the $\gamma-$ray flux obtained in our simulations.

In this work, we constrain the emission of high-energy cosmic messengers from galaxy clusters. 
We employ a detailed numerical framework that combines high-resolution SLOW MHD simulations with Monte Carlo modeling of CR transport in the ICM. 
In addition, we adopt a physically motivated CR injection scheme in which the injection follows the gas density profiles of the clusters. 
In galaxy clusters, regions of high gas density are associated with higher thermal pressure, shock structures, and turbulence, which are physical sites for efficient CR acceleration and injection into the ICM. 
According to our model, the predicted $\gamma$-ray flux lies within the projected sensitivity ranges of CTA and LHAASO, while the predicted neutrino flux is comparable to the expected sensitivities of IceCube-Gen2 and KM3NeT, allowing these observatories to test our predictions.

%

Our present study introduces several methodological improvements compared to our previous works \citep{hussain2021high, hussain2023diffuse, hussain2024neutrinos}, which directly affect the predicted $\gamma-$ray and neutrino backgrounds from galaxy clusters. 
The most important difference in the present study is the adoption of a spatially extended CR injection model. 
The extended injection reduces the concentration of CRs in the central regions, lowering the effective confinement time and hadronic interaction rate, and consequently producing fewer secondaries. 
It also makes the nature of secondary particle production more diffuse than the point-like scenario assumed in previous works, spreading the emission over a larger cluster volume.
%
Another key improvement is the use of the high-resolution SLOW simulations \citep{dolag2023simulating, hernandez2024simulating}, which provide a more realistic description of the ICM compared to the earlier MHD simulation \citep{dolag2005constrained}.
Moreover, our calculation is restricted to clusters within $\sim 500\,\mathrm{Mpc}$ ($z \lesssim 0.1$) distance, whereas previous studies typically integrate over a wider mass and redshift range.

As a result of these improvements particularly the adoption of the extended CR injection model, the tension between the IceCube upper limits and the model assuming a CR energy fraction of $X_{\rm CR} = E_{\rm CR}/E_{\rm cluster} = 0.01$ is \emph{substantially reduced}, even though the same energy fraction is adopted. The more realistic spatial distribution of CR injection is the primary reason for the lower predicted flux, while the updated MHD modeling and focus on local clusters further contribute to this effect, bringing our predictions into better agreement with current observational constraints.

We would like to mention that some parameters may influence our simulations, potentially. For instance, we considered only protons, but clusters can host compact objects like Magnetars or sources such as starburst galaxies that can produce heavy ions. As discussed in section \ref{sec:method}, this could only cause a minor impact on our results.
In addition, the CR composition is quite complex as reported by the Telescope Array \citep{telescope2022cosmic} and the Auger collaborations \citep{aab2017inferences}.
In the MHD simulation, the AGN feedback is not considered, which could influence
the energy budget of CR acceleration processes. Also, the feedback \citep{barai2016kinetic} in the form of powerful jets and outflows injects sufficient energy into the ICM to compensate for radiative cooling, maintaining extremely high temperatures. The feedback process can also change the gas and magnetic field distributions of ICM.
%
%
Galaxy clusters possess very complex structures, a high-resolution MHD simulation including AGN feedback is needed to better constrain the properties of ICM.
Exploring these aspects is one of our plans.

The IGMF is not considered during the propagation of $\gamma$ rays in this work because its strength and structure remain highly uncertain. Current constraints span several orders of magnitude, with upper bounds of $\sim1$~nG from the Planck survey \citep{ade2016planck} and cosmological MHD simulations \citep{dolag2005constrained}, while lower limits in cosmic voids may reach $\sim10^{-17}$ G \citep{acciari2023lower}. Depending on the environment, magnetic fields may vary from $\sim1\mu$G in filaments \citep{bagchi2002evidence} to much weaker values in voids \citep{kotera2008inhomogeneous}. Such variations can affect the development of electromagnetic cascades, as stronger fields deflect the secondary $e^\pm$ pairs and spread the resulting $\gamma$-ray emission over large angular scales, potentially reducing the observable flux from a given cluster.
A systematic exploration of the impact of different IGMF models on the predicted $\gamma$-ray flux is beyond the scope of the present work and will be investigated in a future study.

%

Although our results suggest that galaxy clusters are unlikely to dominate the diffuse high-energy neutrino background, they may still account for a non-negligible fraction of it. 
Constraining their $\gamma-$ray and neutrino emission, therefore, provides direct insight into the efficiency of hadronic processes in large-scale structures and the confinement of CRs in the ICM. 
Upcoming facilities such as LHAASO, CTA, IceCube-Gen2, KM3NeT, and POEMMA \citep{denton2020ultrahigh} will probe this parameter space and test whether galaxy clusters constitute detectable sources of high-energy cosmic messengers.

%

%
%
%
%

\section{Summary}\label{sec:summary}
In this work, we investigate the emission of diffuse $\gamma-$rays and neutrinos from galaxy clusters.
We employ a detailed numerical framework that combines SLOW MHD simulations with Monte Carlo modeling.
We summarize our main findings as follows:

\begin{itemize}
\item 
The SLOW simulations provide the background magnetic field, temperature, and gas distributions of the ICM. In this work, we focus on three massive clusters, namely Virgo, Perseus, and Coma, in the nearby Universe.
The multi-dimensional Monte-Carlo simulations are used to study the propagation of CRs and $\gamma-$rays in the ICM and IGM, considering all relevant interactions and energy-loss mechanisms.
Furthermore, we employed a CR injection model that traces the spatial distribution of gas within clusters, providing a physically motivated approximation. 

\item 
In this work, $\gamma-$rays and neutrinos are produced by large-scale CR interactions within the cluster environment, making the nature of these messengers intrinsically diffusive.

\item 
The diffuse $\gamma-$ray flux predicted by our numerical framework for Virgo, Perseus, and Coma clusters is well below the upper limits reported by the LHAASO collaboration.
Also, the $\gamma-$ray predictions by MAGIC for the central source NGC1275 of the Perseus cluster are slightly higher than our results.

\item 
Furthermore, we estimated the integrated diffuse $\gamma-$ray and neutrino fluxes from the entire population of massive clusters within a $500$~Mpc distance. The predicted neutrino flux is comparable to the existing IceCube upper limits, but the $\gamma-$ray flux is significantly lower than the diffuse flux observed by Fermi-LAT.

\item
$\gamma-$ray emission has been observed from sources such as NGC 1275 and M87, but no significant diffuse $\gamma-$ray emission has yet been detected from the clusters that host them.
The neutrino flux we estimated from the Virgo cluster is comparable to the sensitivity of KM3Net.
Near-term future observations by IceCube-Gen2 and KM3Net might be able to confirm these results.
Similarly, $\gamma-$ray observatories like CTA and LHAASO would be able to observe diffuse emission from galaxy clusters.

\end{itemize}

In summary, our work aligns with the prospects of $\gamma-$ray, neutrino, and CR observatories, making it a timely and valuable study. It further provides important insights into CR acceleration processes in galaxy clusters.



\begin{acknowledgments}
This publication has received funding from the European Union’s Horizon Europe research and innovation program under the Marie Sklodowska-Curie COFUND Postdoctoral Program grant agreement No.101081355-SMASH and from the Republic of Slovenia and the European Union from the European Regional Development Fund.
The simulations in this work were carried out using the VEGA high-performance computing (HPC) system.
We are grateful for the support of the Slovenian National HPC initiative and EuroHPC JU under the project "VEGA - National Supercomputing Centre".
KD acknowledges support by the COMPLEX project from the European Research Council (ERC) under
the European Union’s Horizon 2020 research and innovation program grant agreement ERC-2019-AdG 882679.
The calculations for the hydrodynamical simulation SLOW was carried out at the
Leibniz Supercomputer Center (LRZ) under the project pn68na (CLUES).
\end{acknowledgments}

\bibliography{References_2025_APJ}
\bibliographystyle{aasjournal}

\appendix

\section{CR Injection profiles}
\begin{figure}
\centering
\includegraphics[width=0.45\linewidth]{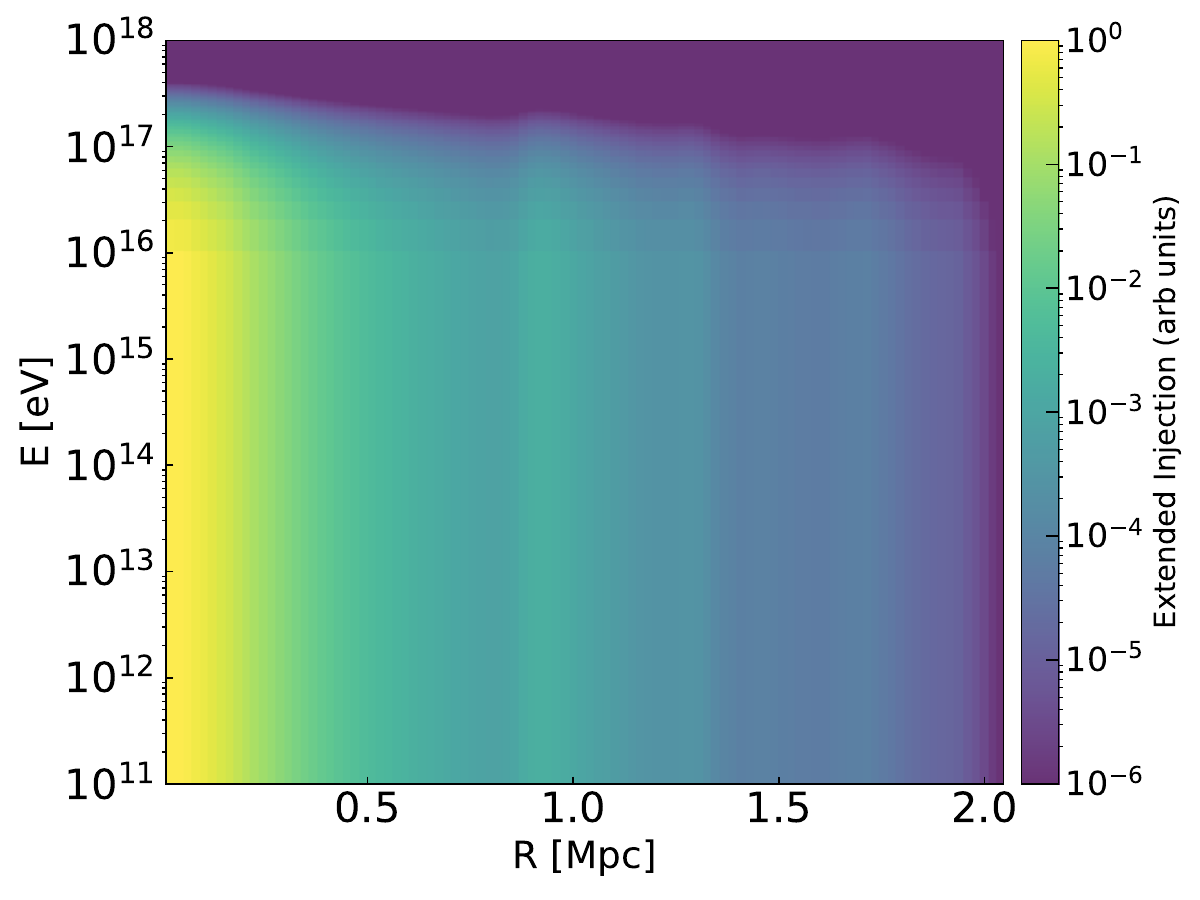}
\hfill
\includegraphics[width=0.45\linewidth]{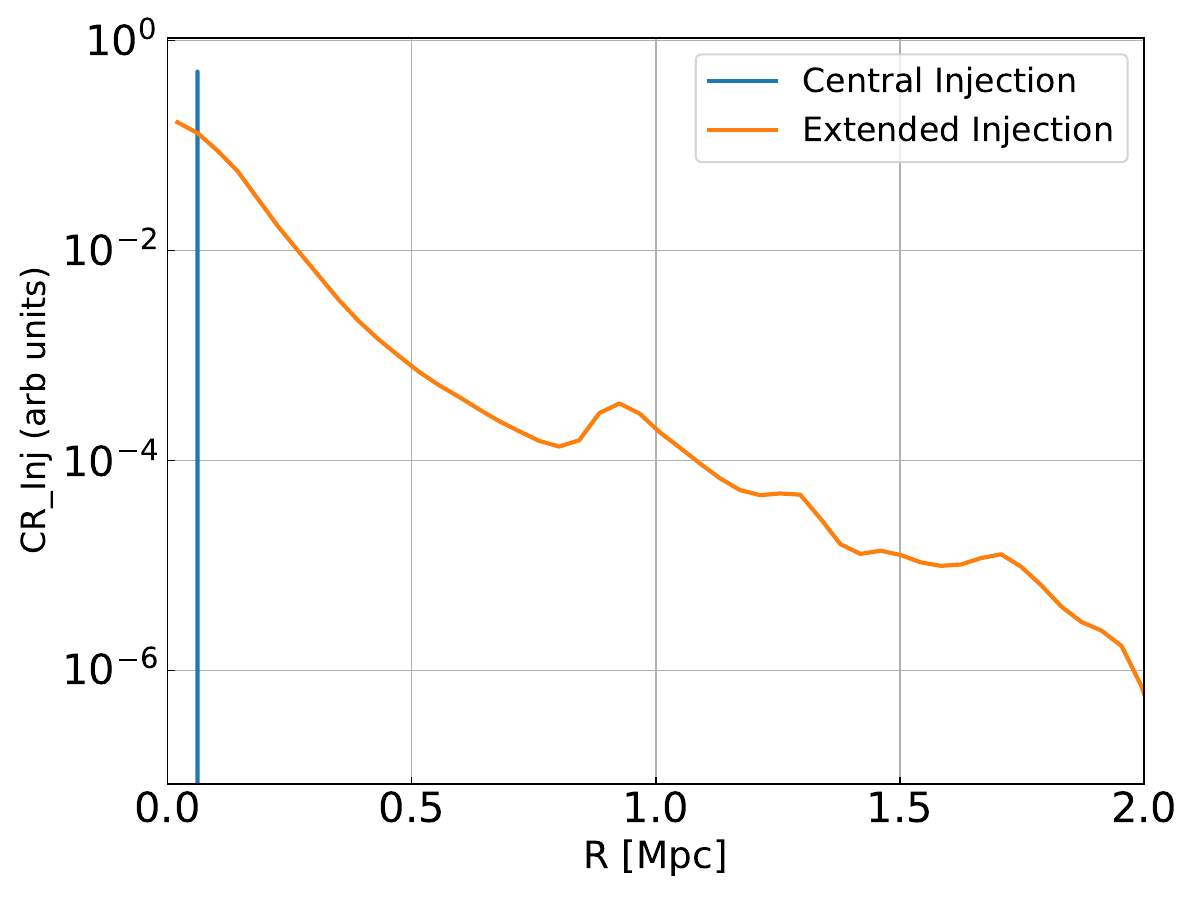}
\caption{Left panel: 2D map of extended injection as a function of radius and CR energy. The color scale shows the injection rate in arbitrary units. The injection follows the volume-averaged profile of gas density of a Perseus-like cluster.
Right panel: Radial profile of CR injection. The central injection is represented by a point source (delta-function) at the cluster core, while the extended injection follows the volume-averaged density profile. Both are plotted in arbitrary units to compare the relative distribution.}
\label{fig:CRInjection}
\end{figure}

The Fig. \ref{fig:CRInjection} represents the CR injection profiles, the central injection used in previous studies \citep{hussain2023diffuse, hussain2021high} and the extended CR injection employed in this work.
%
We show the CR injection in our simulated cluster using two complementary representations. The 2D energy-radius map illustrates how the CR injection varies with radius and energy, highlighting that most of the injection occurs near the cluster center but extends throughout the halo following the gas density profile. To provide a clearer view of the radial dependence, we also plot the 1D radial profiles, where the central injection is modeled as a point source (delta-function), and the extended injection follows the volume-averaged density. The injection rates are shown in arbitrary units, as we focus on the relative distribution and morphology rather than absolute normalization.

\section{pp-Interaction}\label{appen:ppInter}
The pp-interactions and the resulting secondary particle spectra are modeled following \cite{kelner2006energy, kafexhiu2016parametrization}.
The pp-interaction rate is highest at the centers of clusters and gradually decreases toward the outskirts, reflecting the drop in gas density.
The MFP of inelastic pp-interaction can be written as
\begin{equation}\label{eq:ppmfp}
    \lambda_{pp}= \frac{1}{K_{pp}\, \sigma_{pp}\, n_i(r_i)}
\end{equation}
where $K_{pp} = 0.5$ is the inelasticity factor, $n_i(r_i)$ denotes the number
density of protons at a given distance $r_i$ from the centre of the cluster
and $E_p$ is the energy of the protons.
The cross-section $\sigma_{pp}$ is given by
\begin{equation}\label{eq:pp-crosssection}
\sigma_{pp}(E_p)
=
\left[
30.7 - 0.96\,\log\!\left(\frac{E_p}{E_{p,\mathrm{th}}}\right)
+ 0.18\,\log^2\!\left(\frac{E_p}{E_{p,\mathrm{th}}}\right)
\right]
\left[
1 - \left(\frac{E_{p,\mathrm{th}}}{E_p}\right)^{1.9}
\right]^3
\;\mathrm{mb}.
\end{equation}
where $E_p$ is the energy of the proton and $E_{\rm th}$
p is the threshold
kinetic energy $E_{\rm th}$
$p = 2m_\pi + m_\pi /m_p \approx 0.2797 \rm{GeV}$. Eqs. \ref{eq:ppmfp} and \ref{eq:pp-crosssection} are used to calculate $\lambda_{pp}$.

\section{Mass and Redshift dependence of Flux}\label{append:mass-Redshift-depend}
\begin{figure}
\centering
\includegraphics[width=0.45\linewidth]{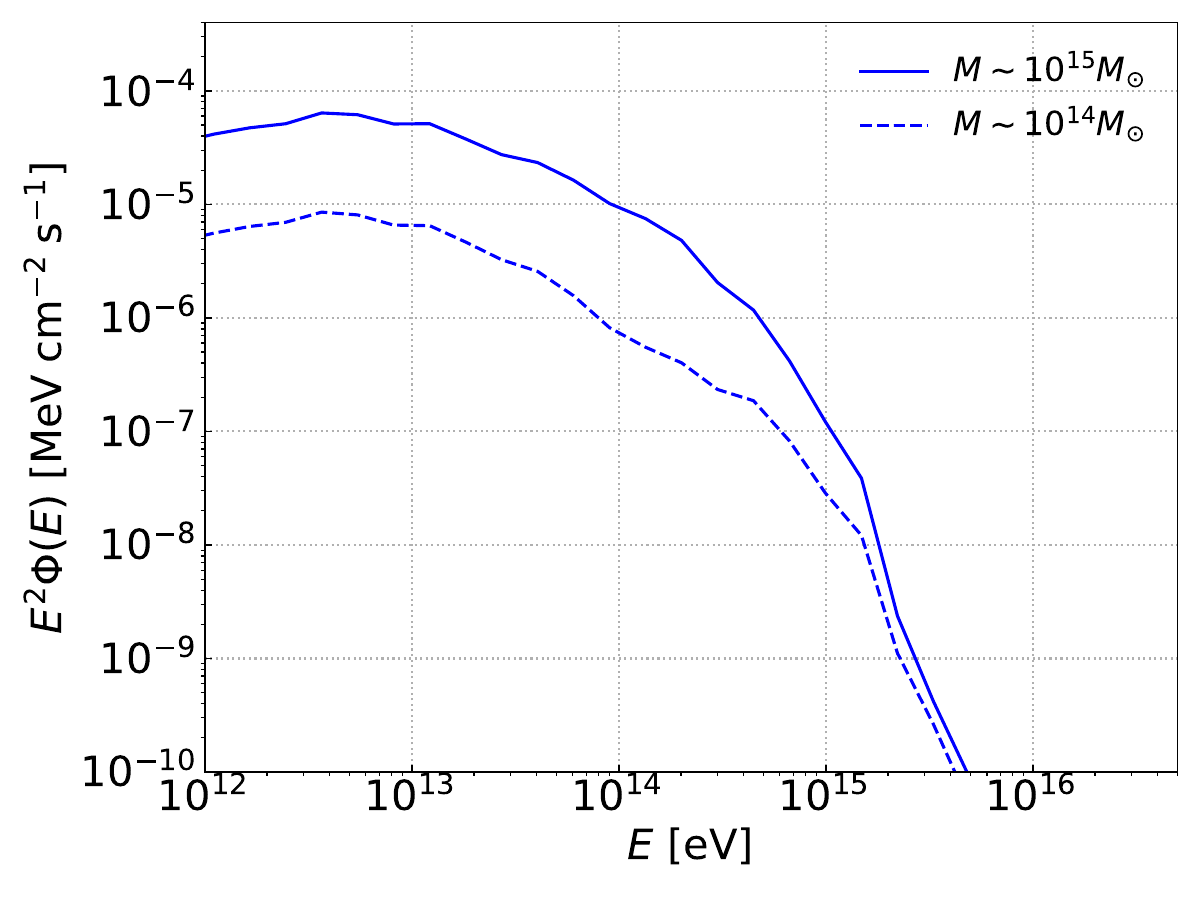}
\hfill
\includegraphics[width=0.45\linewidth]{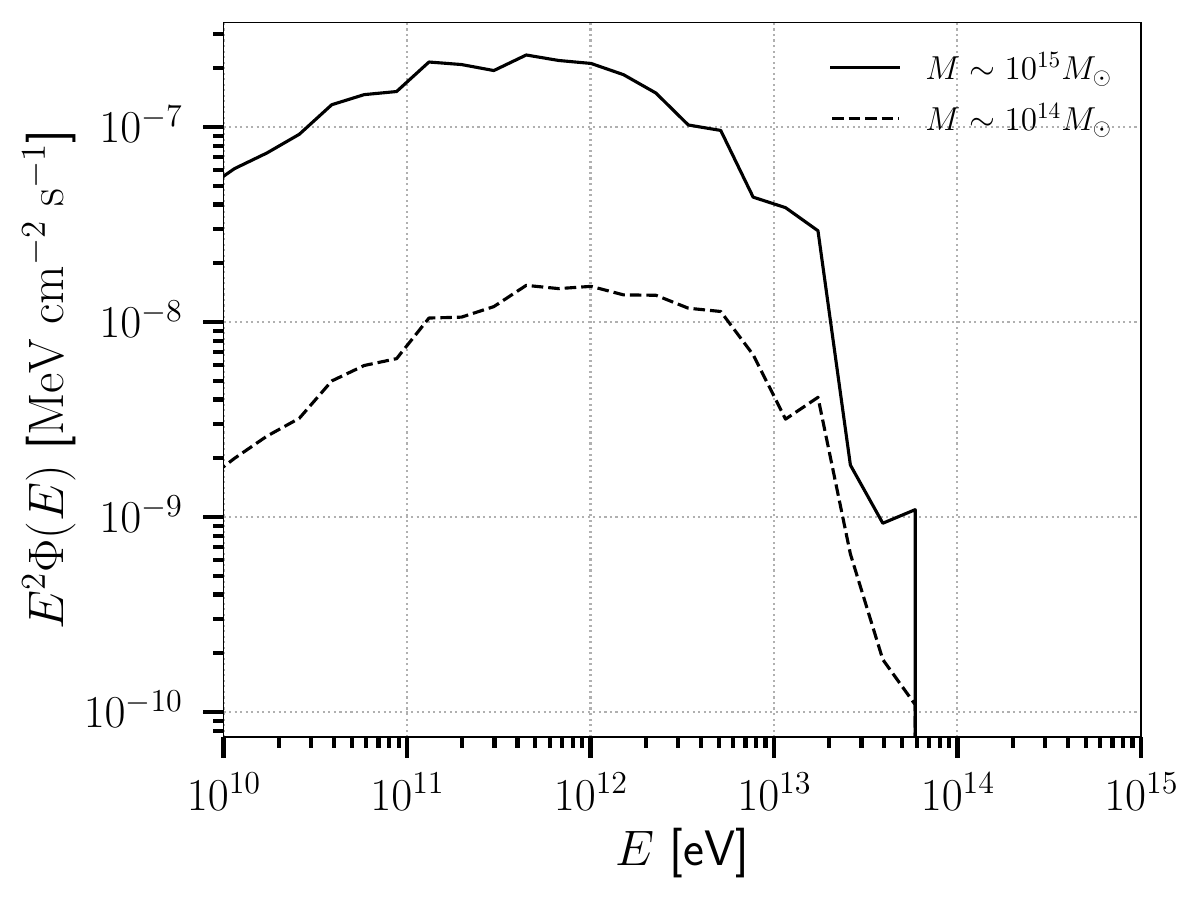}
\hfill
\includegraphics[width=0.45\linewidth]{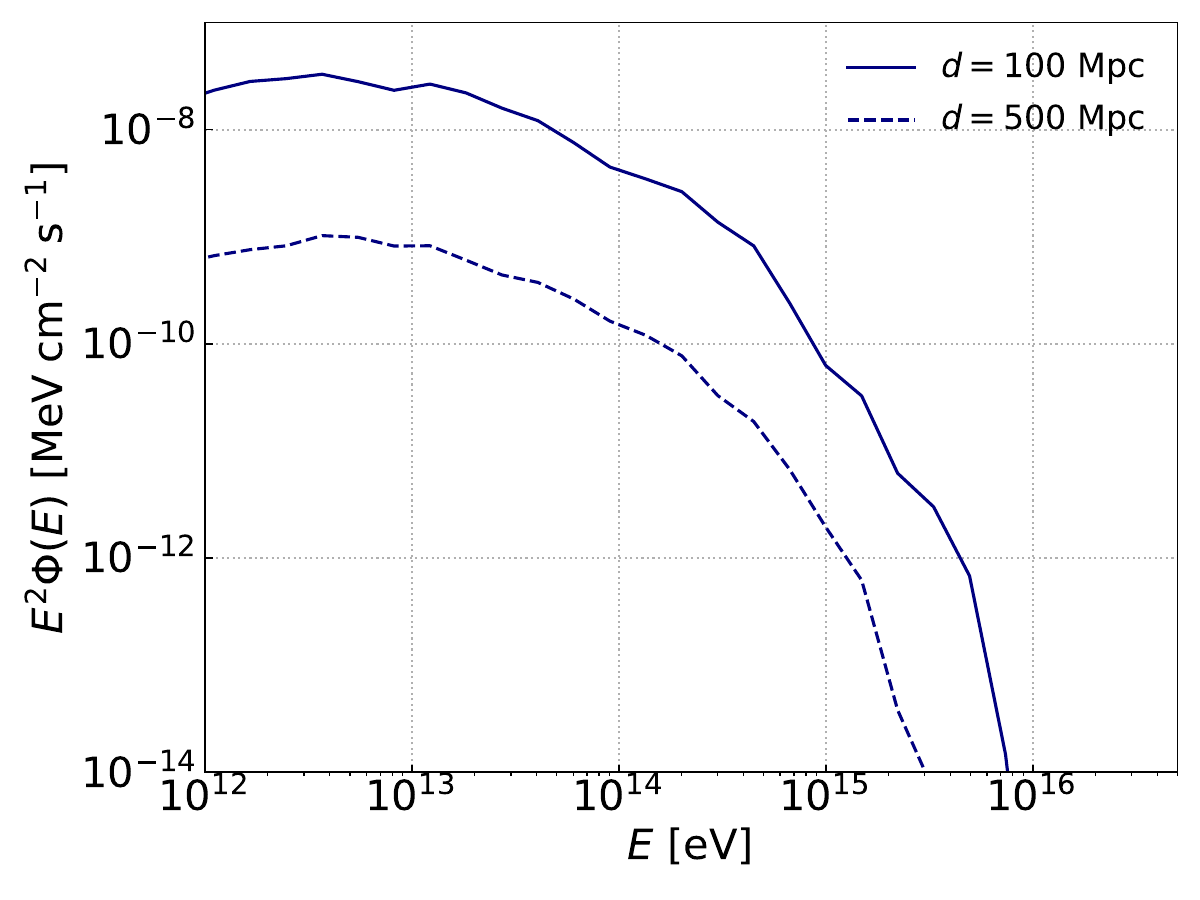}
\hfill
\includegraphics[width=0.45\linewidth]{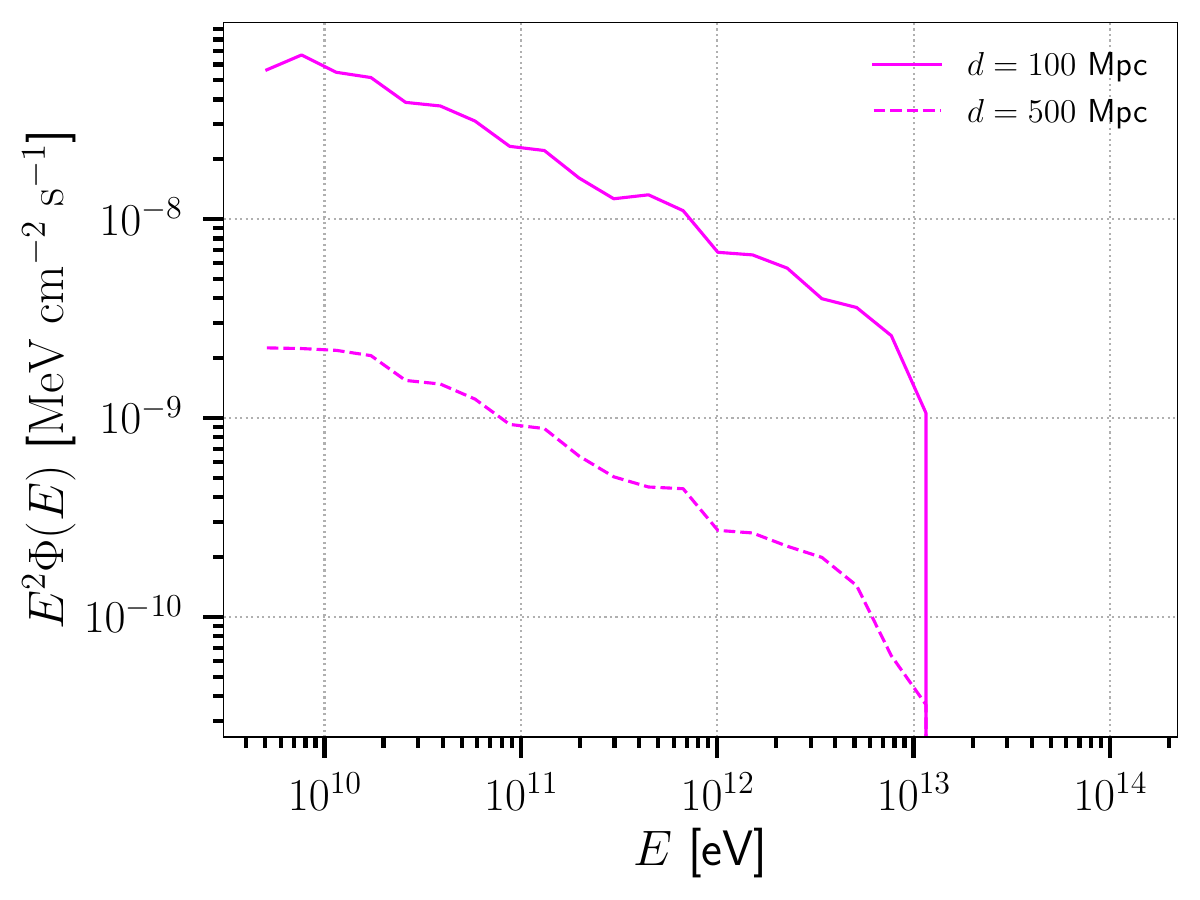}
\caption{
Upper panels represent neutrino and $\gamma-$ray fluxes collected at the edge of two clusters of mass $10^{15}\, M_{\odot} \, \&\, 10^{14}\, M_{\odot}$. Lower panels represent the neutrino and $\gamma-$ray fluxes of two clusters of similar mass ($\sim10^{15}\, M_{\odot}$) but at different distances ($d=100\, \rm{Mpc}\, \& \, d=500 \, \rm{Mpc}$).
}\label{fig:flux_ZMdepend}
\end{figure}

To demonstrate the dependence of the predicted multi-messenger signals on cluster properties, we performed Monte Carlo simulations for a representative set of clusters based on the physical conditions derived from the SLOW simulations. The resulting $\gamma$-ray and neutrino spectra were examined for clusters with different masses and distances. As examples, we present results for clusters with masses $10^{15}\,M_\odot$ and $10^{14}\,M_\odot$ to demonstrate the mass dependence, and for clusters located at distances (comoving) of $100$~Mpc and $500$~Mpc to illustrate the redshift dependence. These examples highlight how the predicted emissions vary with the underlying cluster parameters.
In Fig. \ref{fig:flux_ZMdepend}, we presented the $\gamma-$ray and neutrino fluxes of clusters of different masses and distances (comoving).

\end{document}